\documentclass[floatfix,superscriptaddress,a4paper,showkeys,nofootinbib,prc]{revtex4-2}
\usepackage[colorlinks=true,linktocpage=true,linkcolor=blue,citecolor=blue,allcolors=blue]{hyperref}

\usepackage{epsfig}
\usepackage{latexsym}
\usepackage[utf8]{inputenc}
\usepackage{xspace}
\usepackage{indentfirst}
\usepackage{enumitem}
\usepackage{placeins}
\usepackage[table,xcdraw]{xcolor}
\usepackage{setspace}
\usepackage{lipsum}
\usepackage{multirow}
\usepackage{comment}

\usepackage{graphicx}

\usepackage{amsmath}
\usepackage{amssymb}
\usepackage[english]{babel}

\usepackage{url}
\graphicspath{{figs/}}

\newcommand{\mean}[1]{\langle #1 \rangle}
\newcommand{\eq}[1]{\begin{eqnarray} #1 \end{eqnarray}}

\newcommand{\be}{\begin{equation}}
\newcommand{\ee}{\end{equation}}

\begin{document}

\title{Correlations of Conserved Quantities at Finite Baryon Density}
\author{Oleh Savchuk}\thanks{Corresponding author}
\email{savchuk@frib.msu.edu} 
\affiliation{Facility for Rare Isotope Beams, Michigan State University, East Lansing, MI 48824 USA}
\affiliation{Bogolyubov Institute for Theoretical Physics, 03680 Kyiv, Ukraine}
\author{Scott Pratt}
\affiliation{Department of Physics and Astronomy, Michigan State University, East Lansing, MI 48824 USA}
\affiliation{Facility for Rare Isotope Beams, Michigan State University, East Lansing, MI 48824 USA}
\date{\today}
\begin{abstract}
Correlations involving the seven conserved quantities, namely energy, baryon number, electric charge, strangeness, and the three components of momentum, give rise to correlations in heavy-ion collisions. Through the utilization of a simple one-dimensional hydrodynamic model, we calculate the evolution of the entire $7\times 7$ matrix of correlations as a function of relative spatial rapidity. This comprehensive analysis accounts for finite baryon density, which results in off-diagonal correlations between the charge-related quantities and the energy-momentum quantities. These correlations in coordinate space are subsequently transformed into correlations in momentum space using statistical weighting. The entire matrix of correlations is revealed to be highly sensitive to the equation of state (EoS), viscosity, and diffusivity.
\end{abstract}

\maketitle

\section{Introduction}
The properties of QCD matter have been a major focus of the nuclear physics community~\cite{Busza:2018rrf,Bzdak:2019pkr}. 
Nucleus-nucleus (A+A) collisions provide the majority of experimental information about the bulk properties of the matter reflected in its phase diagram. Particularly, the existence phase transitions (PTs) related to the change of hadronic to quark and gluon degrees of freedom have become a focal point of the field. Recently, the properties of cold and dense nuclear matter reached in low energy collisions has become increasingly relevant in the context of binary neutron star mergers. Observations related to the radii of neutron stars point to a remarkably stiff equation of state for baryon densities reaching a few times nuclear saturation density, with the speed of sound possibly exceeding half the speed of light \cite{Dietrich_2020,PhysRevC.100.055801,
Hanauske2020a,Hanauske2020b,Schaffner2018,
PhysRevD.104.063003, Drischler_2021,Huth_2022,https://doi.org/10.48550/arxiv.2208.00994,Sorensen:2023zkk,PhysRevLett.105.161102,PhysRevC.85.032801, PhysRevC.89.025806}. Fluctuation and correlation observables have always played a prominent role in representing the bulk properties of matter, particularly near phase transitions or near critical points~\cite{Steinheimer:2012gc,Steinheimer:2013glaa,Savchuk:2022msa,Vovchenko:2020tsr}. Even for featureless equations of state fluctuations represent fundamental physical quantities. For example, the speed of sound can be expressed in terms of fluctuation observables~\cite{Sorensen:2022odd}.

Unfortunately, the matter created in A-A collisions evolves and disassociates in a short time. The ephemeral nature of the produced matter impedes the opportunity for longer-range correlations to equilibrate. Thus, in the context of heavy-ion collisions, the dynamical evolution of correlations must be understood. Whereas in a macroscopic system with infinite time to equilibrate, the charge within any finite volume fluctuates, in heavy-ion collisions one needs to carefully specify the volume over which the fluctuation is being considered. For conserved charges, there is zero fluctuation in the limit that one considers the entire collision volume.

To understand correlations and fluctuations from a dynamic perspective, one must understand how correlations are seeded, and how they spread with time. For conserved charges, such as baryon or electric charge, this  is typically quantified through charge balance correlations \cite{Ling:2013ksb,Bass:2000az}. In coordinate space such correlations can be expressed as 
\begin{eqnarray}
C(\vec{r}_1,\vec{r}_2,t)&=&\langle[\rho(\vec{r}_1,t)-\bar{\rho}(\vec{r}_1,t)][\rho(\vec{r}_2,t)-\bar{\rho}(\vec{r}_2,t)]\rangle.
\end{eqnarray}
If the correlations appear equilibrated at short relative distance, one can divide the correlation into two pieces,
\begin{eqnarray}
C(\vec{r}_1,\vec{r}_2,t)&=&\chi(\vec{r}_1)\delta(\vec{r}_1-\vec{r}_{2})+C_B(\vec{r}_1,\vec{r}_2,t).
\end{eqnarray}
Here, the delta function is not meant literally, but only to represent some function of short range that integrates to unity. If a system is equilibrated, and if those correlations are not longer-range, the strength of the local correlation is set by the charge susceptibility, which is a fundamental bulk property of the matter. The second term, the balancing correlation, would spread over the entire available volume if given the opportunity. However, its spread is constrained by the amount of time elapsed since the correlations were seeded. The integrated strength of the balancing correlation must reflect the fact that for the entire volume, charge does not fluctuate, and
\begin{eqnarray}
\int d^3r~C_B(\vec{R},\vec{R}+\vec{r},t)&=&-\chi(\vec{R},t).
\end{eqnarray}
If the system has translational invariance, the $\vec{R}$ dependence can be ignored. At the end of the collision, one can measure the balancing correlation, i.e. it is the correlations of all particles with those besides themselves. At high energy, where the average charge densities are near zero, $C_B(\vec{r},t)$ typically spreads diffusively. Local charge conservation demands that the source function for the balance correlations, in the matter frame, is given by
\begin{eqnarray}
S(\vec{R},t)&=&-(\partial_{t}+\nabla\cdot\vec{v})\chi(\vec{R},t).
\end{eqnarray}
Thus, if one measures the correlations at latter times, and if one understands how charges diffuse with time, one can infer the source function for the correlations. One can then infer, not only the susceptibility at a specific time, but the evolution of the susceptibility over time. Hydrodynamic calculations provide temperatures, flow velocities and densities as a function of time. This then provides the susceptibilities, which are functions of density and temperature, as a function of time, which then determines the source functions for $C_{B}$, and if one were to know the diffusivity, one can then calculate the evolution of $C_{B}$. Indeed, such calculations have been performed, overlaid on a three-dimensional hydrodynamic background, using the equation of state, charge susceptibility, and diffusivity from lattice calculations \cite{Pratt:2015jsa,Pratt:2016lol,Pratt:2018ebf,Pratt:2021xvg}. 

In heavy-ion collisions, measurement is confined to recording the momenta of final-state particles. Fortunately, collisions are highly explosive, and the large flows provide the means to map correlations in relative velocity to those in coordinate space, albeit with some smearing due to mainly thermal motion. The corresponding correlations in relative velocity have been measured, and are known as charge-balance functions. Such analyses, which can be binned by any measure of the relative velocity, such as the relative momentum or relative azimuthal angle, have proven remarkably adept at constraining both the evolution of the charge susceptibilities as a function of time, and the diffusivity of QCD matter. Remarkably, the analyses support the notion that matter chemically equilibrates early in the collision, and that baryon number, strangeness and electric charge diffuse at rates calculated in lattice simulations \cite{Aarts:2014nba}.

The first goal of this paper is to understand how the above paradigm can be extended to include other conserved quantities, i.e. energy and momentum. The second goal is to understand how at non-zero average charge density, which is increasingly the case for lower-energy collisions, energy and momentum correlations mix with those indexed by charge. Whereas the first goal was considered in \cite{Pratt:2016lol}, the effects of non-zero density were ignored. One can define a $7\times 7$ susceptibility matrix, involving energy, three components of momentum, and the three charges, baryon number, electric charge, and strangeness. That matrix has cross terms between energy and the three charge densities. For example, the fluctuation between energy and baryon density, $\langle\delta E\delta B\rangle$, is non-zero if there is a finite net baryon density. To make matters more complicated, energy and momentum correlations do not spread diffusively, but due to viscous hydrodynamics. Energy and momentum correlations then mix through the hydrodynamic evolution equations, regardless of whether the net baryon density is zero. The next goal is to show how correlations involving energy and momentum in coordinate space can be projected onto final-state hadrons at the decoupling by providing a thermal sampling of particles, with each particle accompanied by an additional weight factor. Finally, as a fourth goal, we wish to see how the corresponding $7\times 7$ matrix of correlation functions might be sensitive to variations of the equation of state. 

The model to be applied here is simple, longitudinally-boost-invariant hydrodynamics without transverse flow. Hereafter, this will be referred to as the Bjorken limit \cite{Bjorken:1982qr}. The neglect of transverse flow is not warranted, at any energy, and boost invariance is poorly justified at lower energies. However, this simple picture should be sufficient for understanding the issues mentioned above. Correlations involving energy and momentum are likely to be strongly affected by inhomogeneities of the initial state, and by jets. Even crude estimates of the strength of correlations due to local charge/energy/momentum conservation are valuable in that one can then ascertain whether further, more realistic, modeling is warranted, or if such correlations are so small that they are overwhelmed by the competing effects related to the initial state and jets.

The next section of the paper focuses on the underlying theory. It presents the generalized $7\times 7$ susceptibilities, and the expressions for then evolving the $7\times 7$ matrix of balancing correlations. The evolution is based on small perturbations of the uniform hydrodynamic background, which spread both diffusively and hydrodynamically. Green's functions are pursued as a means by which to evolve the correlations from the sources. Expressions for the evolution on the simple boost-invariant background are presented. The final subsection of the theory section shows how one can project the spatial correlations onto a simulation of outgoing particles, which can then be used to calculate how such spatial correlations might drive the asymptotic final-state correlations measured in momentum space. The final-state correlations can be binned by any measure of relative momentum, such as relative rapidity or azimuthal angle, and might be indexed by such quantities as transverse energy, transverse momentum, or charge. 

In Sec. \ref{sec:results}, after showing an illustrative example of Green's functions, final-state correlations are presented for three different equations of state. Correlations are presented both in coordinate space, as a function of relative spatial rapidity, and momentum space as a function of relative rapidity. Implications of the results are then summarized in Sec. \ref{sec:conclusions}.

\section{Theoretical Foundation}\label{sec:theory}

This section consists of three parts. First, the connection between the $7\times 7$ susceptibilities and the correlation functions are elucidated. The source function, which is driven by the evolution of the susceptibility, is connected to the final-state correlation through Green's functions which describe the response of the medium to small perturbations of the seven quantities. The following sub-section shows how Green's functions are calculated on the background of a boost-invariant hydrodynamics description lacking transverse flow. Finally, a method is presented for projecting the correlations onto the momenta of particles in a hadron gas.

\subsection{General Theory of Conservation-Driven Correlations}\label{subsec:generaltheory}

Susceptibilities are related to fluctuations, as calculated in the Grand Canonical Ensemble. In general, one can weight a partition function according to any conserved quantity,
\begin{eqnarray}
Z&=&{\rm Tr}~\exp\{-\beta_iQ_i\}.
\end{eqnarray}
Here, Lagrange multipliers, $\beta_i$, are associated with each conserved quantity $Q_i$. The seven quantities, $Q_i$, are $E,P_x,P_y,P_z,B,Q,S$, the energy, momenta, baryon number, electric charge and strangeness. For some large volume $\Omega$, one can define fluctuations as
\begin{eqnarray}
\frac{1}{\Omega}\langle\delta Q_i\delta Q_j\rangle&=&\frac{1}{\Omega}\frac{\partial^2}{\partial\beta_i\partial\beta_j}\ln(Z).
\end{eqnarray}
The Lagrange multiplier related to the energy is the inverse temperature, whereas for the three charges they are $-\mu_i/T$, where $\mu_i$ is the chemical potential and $T$ is the temperature. For the momenta, the Lagrange multipliers can be related to $u_i/T$, where $u_i$ are the flow velocities. Here, the term susceptibilities is used in a most general sense. The usual, but not always, definition of charge susceptibilities is to define them as the derivative of the charge densities with respect to the chemical potentials. That definition differs from what is used here by a factor of $T$. In the usage here, all the ``susceptibilities'' correspond to the fluctuations without additional factors,
\begin{eqnarray}
\chi_{ij}&=&\frac{1}{\Omega}\langle\delta Q_i\delta Q_j\rangle.
\end{eqnarray}
For the energy, the susceptibility, $\langle\delta E\delta E\rangle/\Omega$, is $T^2$ multiplied by the specific heat. The $3\times 3$ subset of $\chi_{ij}$ that refer to baryon number, electric charge and strangeness, are the usual charge fluctuations. At finite number baryon density, the correlations between the energy and the three charges becomes non-zero. These cross terms have not been considered as extensively as correlations involving only the three charges or only the energy. Finally, one can have correlations involving the three momentum components. These have no cross terms, and at equilibrium
\begin{eqnarray}
\frac{1}{\Omega}\langle \delta P_i\delta P_j\rangle&=&(P+\epsilon)T\delta_{ij}.
\end{eqnarray}

In equilibrium, the correlations integrate to the susceptibilities. If $\rho_i$ are the corresponding densities,
\begin{eqnarray}
C_{ij}(\vec{r}_i,\vec{r}_j)&=&\langle\delta\rho_i(\vec{r}_i)\delta\rho_j(\vec{r}_j)\rangle\\
\nonumber
&=&\chi_{ij}\delta(\vec{r}_i-\vec{r}_j).
\end{eqnarray}
In a heavy ion collision, all the quantities are conserved, so for any operator $A$,
\begin{eqnarray}\label{cons-rule}
\langle A\delta Q_j\rangle&=&0,\\
\nonumber
\int d^3r_j\langle A\delta\rho_j(\vec{r}_j)\rangle&=&0,
\end{eqnarray}
and for any conserved density
\begin{eqnarray}
\int d^3r'~C_{ij}(\vec{r},\vec{r}')&=&0.
\end{eqnarray}
The correlation can be divided into a local, equilibrated, part and a balancing part,
\begin{eqnarray}
C_{ij}(\vec{r}_i,\vec{r}_j)&=&\chi_{ij}\delta(\vec{r}_i-\vec{r}_j)+C_{B,ij}(\vec{r}_i-\vec{r}_j),\\
\nonumber
\int d^3r_jC_{B,ij}(\vec{r}_i-\vec{r}_j)&=&-\chi_{ij}(\vec{r}_i).
\end{eqnarray}
If the local part were not equilibrated, the equivalence would still hold true, but in that case $\chi_{ij}$ would not be given by the equilibrated susceptibility. Instead, it might be some dynamic quantity that would relax toward the equilibrated value. 

As the system cools into a gaseous state, correlations between particles disappear. In that limit, the equilibrated local correlation is only that within a single particle, and the susceptibilities become
\begin{eqnarray}
\chi_{ij}&=&\frac{1}{\Omega}\sum_{a\in\Omega} Q_{ai}Q_{aj},
\end{eqnarray}
where $Q_{ai}$ is the charge of type $i$ on particle $a$. In terms of the phase space density,
\begin{eqnarray}
\chi_{ij}(\vec{r})&=&\frac{1}{(2\pi)^3}\sum_h\int d^3p~f_h(\vec{p},\vec{r}) Q_{i}(h,\vec{p})Q_{j}(h,\vec{p}),
\end{eqnarray}
where the sum extends over hadron species. For example, if $i$ refers to the momentum component $p_x$, $Q_i(\vec{p})=p_x$ and if $i$ refers to strangeness, $Q_i$ is simply the strangeness of species $h$. Thus, at decoupling, the balancing correlation, $C_{B,ij}$, covers all correlations between different particles, assuming the matter is a non-interacting gas at the time of decoupling.

In the absence of a source function $\langle\delta\rho_i(x_1)\delta\rho_j(x_2)\rangle$ evolves as $\langle\delta\rho_i(x_1)\rangle\langle\delta\rho_j(x_2)\rangle$ with each $\delta\rho$ behaving independently in terms of the space time coordinates $x_1$ and $x_2$. To satisfy the sum rule, one can then express the correlation as
\begin{eqnarray}
\langle\delta j^\mu_i(x_1)\delta j^\nu_j(x_2)\rangle&=&
\int d^4X~G^{\mu\mu'}_{ii'}(X,x_1)G^{\nu\nu'}_{jj'}(X,x_2)u_{\mu'}(X)S_{i'j'}(X)u_{\nu'}(X).
\end{eqnarray}
Here, the formalism has been expressed with the charge density replaced by the four-current. This enables one to project the correlations through a hypersurface that is not necessarily at constant Euclidean time. For baryon, electric and strangeness currents, the quantities $j_i^\mu$ are simply the usual three charge currents. The energy and momentum current densities are elements of the stress-energy tensor, $T^{0\mu}$ and $T^{k\mu}$, respectively, where $k$ denotes the $k^{\rm th}$ component of the momentum density. The Green's functions satisfy the normalization conditions,
\begin{eqnarray}
\int d\Omega_\mu G^{\mu\mu'}_{ii'}(X,x)u_{\mu'}(X)=\delta_{ii'},
\end{eqnarray}
where $x$ refers to the space-time point along the hyper-surface  where the matter disassociates and $d\Omega_\mu$ is the differential volume in the reference frame where the disassociation is locally simultaneous.  The extent of this hyper surface is fully in the absolute future relative to $X$, and the integral covers the entire hyper volume. If one defines a step function, such that inside the hyper volume the value is unity, and outside it is zero, the differential volume element can be defined as
\begin{eqnarray}
d\Omega_\mu=d^4y~\partial_\mu\Theta(C(y)),
\end{eqnarray}
where $C(y)$ is positive outside the surface and negative inside. For example, defining $C(y)=T_0-T$, defines a hyper-surface separating the regions with temperature above and below $T_0$. In this study, where a simple Bjorken expansion is considered, space-like hyper-surfaces, like one would see with evaporation from a surface, are not encountered.

The source functions are given by the rate of change of the susceptibilities. If the collective velocity is noted by $u^\mu(x)$,
\begin{eqnarray}
S_{ij}(X)&=&[u\cdot\partial+(\partial\cdot u)]\chi_{ij}(X).
\end{eqnarray}
In a small hydrodynamic volume $\delta V$, the rate of change of the volume is $\delta V\partial\cdot u$, and the source function is $(1/\delta V)(u\cdot\partial)(\delta V\chi_{ij})$. Thus, if $\chi_{ij}$ falls inversely with the volume there is no source. This would happen for baryon, electric and strange charges for the isentropic expansion of a non-interacting gas. If entropy is conserved, one can identify the source function with 
\begin{eqnarray}
S_{ij}(x)\approx s(u\cdot\partial)\frac{\chi_{ij}}{s}.
\end{eqnarray}
where $s$ is the entropy density, with this becoming exact for isentropic expansions. For non-zero charge densities, one can similarly replace $s$ with the baryon, electric or strange densities, although this would neglect charge diffusion.

The principal challenge in calculating correlations is in finding the Green's functions. At each point in spacetime, $X$, one considers a small perturbation of each of the seven quantities. One then writes equations of motion for the perturbation. These equations mix the various quantities, i.e. $G^\mu_{ii'}(X,x)$ is diagonal for times $x_0$ immediately following $X_0$ but off-diagonal components ensue thereafter. If the average charge densities were all zero, the Green's functions for energy and momentum do not mix with those for baryon number, electric charge or strangeness. The Green's functions for charge then simply require solving the diffusion equation for an initial charge perturbation at $X$. If the diffusivity matrix is diagonal, as one would expect for an idealized quark-gluon plasma, the charge evolutions would not mix the three charges if the were expressed in the $u,d,s$ basis. For the energy and momentum perturbations, one must address the evolution with hydrodynamics. Hydrodynamics mixes the energy and momentum components, and if the average charge densities are non-zero, as is the case for lower-energy heavy-ion collisions, the Green's function also develops off-diagonal components between the energy/momentum and charge components.


\subsection{Hydrodynamic Response}\label{subsec:hydroresponse}
The Green's function represents the response of the medium to a small localized fluctuation. In this study, this is calculated by assigning a small perturbation of charge type $i$ to a specfic point in space-time. The perturbation integrates to unity, but is treated in the linear approximation. Its spatial extent is set to a small, but non-zero, value. In the limit that the extent is zero, the initial perturbation would be a delta function, but that would preclude evolution equations involving spatial derivatives. Here, we first review the treatment of linearized hydrodynamic fluctuations.

The stress-energy tensor of a viscous fluid in the Landau frame is represented as follows~\cite{Romatschke:2010}:
\eq{
T^{\mu\nu} = (\varepsilon + P(\varepsilon,\rho_B,\rho_Q,\rho_S))u^\mu u^{\nu} - g^{\mu\nu}P + T_{\eta}^{\mu\nu}.
}
In this equation, $\varepsilon$ and $\rho$ stand for energy and particle densities, respectively. In this study, the shear contribution $T^{\mu\nu}_{\eta}$ is assigned according to the Navier-Stokes equation,
\eq{
T^{\mu\nu}_{\eta} = -\eta\left(\nabla^{\mu}u^{\nu}+\nabla^{\nu}u^{\mu}-\frac{2}{3}\Delta^{\mu\nu}\partial\cdot u\right),
}
where $\eta$ is the shear viscosity coefficient, and $\Delta^{\mu\nu}=u^{\mu}u^{\nu}-g^{\mu\nu}$ acts as a projector eliminating the collective velocity $u^\mu$, and the covariant derivative $\nabla^{\mu}=\Delta^{\mu\nu}\partial_\nu$ represents the spatial derivatives in the frame of the fluid. If the charge number is allowed to diffuse, the charge current can be expressed as:
\eq{
j^{\mu}_i = \rho_i u^{\mu} - D_{ij}\Delta^{\mu\nu}\partial_{\nu}\rho_j.
}
In this equation, $D$ represents the diffusion matrix. The hydrodynamic equations are equivalent to energy-momentum conservation:
\begin{eqnarray}
\partial_{\nu}T^{\mu\nu} &=& 0,\\
\nonumber
\partial_{\nu}j^{\nu} &=& 0.
\end{eqnarray}

Assuming that the solution for $\rho(t,\vec{r}),\varepsilon(t,\vec{r})$ and $u^{\mu}(t,\vec{r})$ can be expanded into a series around the given solution, with respect to the perturbations $\delta \rho,\delta \varepsilon ,\delta u^{\mu}$, our focus will mainly be on $\delta T^{\mu\nu}$, which is linear in these perturbations. These perturbations should adhere to the following equations:
\begin{eqnarray}\label{hydro-linear1}
\partial_{\nu}\delta j^{\nu} &=&0,\\
\label{hydro-linear2}
\partial_{\nu}\delta T^{\mu\nu} &=& 0.
\end{eqnarray}

To generate the Green's function, one considers a small very localized perturbation at space-time point $X$ that evolves over time. One can assign a small Gaussian perturbation. If the initial perturbation is at space-time point $x$, 
\begin{eqnarray}
\delta j^\mu_{i'}(X)&=& \frac{1}{(2\pi\sigma^2)^{3/2}}e^{-|Y^2|/2\sigma^2},\\
\nonumber
Y^\mu&=&X^\mu-(u\cdot X)u^\mu.
\end{eqnarray}
Here, $Y$ is the same as $X$, but with the temporal component, as defined in the fluid frame, projected away. The width $\sigma$ should be chosen as small as possible while still enabling differentiation. In the limit of $\sigma\rightarrow 0$ the initial perturbation becomes a delta function in coordinate space. For the initial perturbation, one then solves for the perturbation at a space-time point with time $x$, and the Green's function is
\begin{eqnarray}
G^\mu(X,x)_{ii'}&=&\delta j^\mu_i(x).
\end{eqnarray}

Given the existence of dissipation, all solutions should asymptotically approach a uniform solution. This is particularly relevant for thermodynamic fluctuations in baryon number, momentum, or energy. Following the theoretical framework outlined in \cite{Pratt:2016lol}, we aim to investigate the correlations of conserved charges at freeze-out, specifically examining the impact of non-diagonal susceptibilities in the presence of finite baryon density.

\subsection{Bjorken Expansion as a Background}
To most simply illustrate how correlations evolve during the hydrodynamic stage of strongly interacting matter, a boost-invariant Bjorken solution is chosen as the background hydrodynamic model. In this case, the evolution is simplified by transforming the time $t$ and longitudinal space-coordinate $z$ to what is known as the proper time $\tau$ and the spatial rapidty, $\eta$:
\eq{
\quad t = \tau \cosh\eta, \quad z = \tau \sinh\eta.
}
Energy and $B,Q,S$ density is uniform through the system and its time evolution is given by:
\eq{
\partial_{\tau}\varepsilon=-\frac{\varepsilon + P}{\tau} + \frac{4\eta_s}{3\tau^2},
}
\eq{
\rho=\rho_0\frac{\tau_0}{\tau},
}
where $\tau_0, \rho_0$ represents the starting time and density. The four velocity is
\eq{u^0=\cosh\eta,\,u^z=\sinh\eta.}
The system of linear response equations is given by:
\eq{\label{eq-e}
\partial_{\tau}\delta\varepsilon &=& -\frac{1}{\tau}\delta\left(\varepsilon+P\right) - \partial_{\eta}\frac{\varepsilon+P-\frac{8\eta_s}{3\tau}}{\tau}\delta u^{\eta},
}

\eq{\label{eq-pz}
\partial_{\tau}\left(\varepsilon+P-\frac{4\eta_s}{3\tau}\right)\delta u^{\eta} &=& -\frac{2}{\tau}\left(\varepsilon+P-\frac{4\eta_s}{3\tau}\right)\delta u^{\eta} - c_{\rho}^2\frac{\partial_{\eta}\delta\varepsilon}{\tau} - \partial_{\rho}P\frac{\partial_{\eta}\delta\rho}{\tau} + \frac{4\eta_s\partial_{\eta}^2\delta u^\eta}{3\tau^2},
}

\eq{\label{eq-rho}
\partial_{\tau}\delta\rho &=& -\frac{\delta \rho}{\tau} + \frac{\rho+D\partial_{\tau}\rho}{\tau}\partial_{\eta}\delta u^{\eta} + \frac{D}{\tau^2}\partial_{\eta}^2\delta\rho,
}

\eq{\label{eq-px}
\partial_{\tau}\left(\varepsilon+P-\frac{4\eta_s}{3\tau}\right)\delta u^{x,y} &=& -\frac{2}{\tau}\left(\varepsilon+P-\frac{4\eta_s}{3\tau}\right)\delta u^{x,y} + \frac{4\eta_s\partial_{\eta}^2\delta u^{x,y}}{3\tau^2},}
where $\delta u^{\eta}=\frac{\delta u^z}{u^0}$ is introduced in order to have boost-invariant system of equations (uniform in $\eta$).
These equations describe the hydrodynamic response to some perturbation given by the stochastic current $j_{\varepsilon},\vec{j}_{\vec{P}},j_B,j_Q,j_S$. This system also conserves laboratory frame energy and momentum:

\eq{
\partial_t T^{tt}+\partial_z T^{tz} &=& \left(\partial_{\tau}+\frac{1}{\tau}\right)(\cosh\,\eta T^{tt}-\sinh\eta\,T^{tz}) \\
\nonumber
&\quad& + \frac{1}{\tau}\partial_{\eta}(-\sinh\eta\,T^{tt}+\cosh\eta\,T^{tz}) = 0,\\
\nonumber
\partial_t T^{tz}+\partial_z T^{zz} &=& \left(\partial_{\tau}+\frac{1}{\tau}\right)(\cosh\,\eta T^{tz}-\sinh\eta\,T^{zz}) \\
\nonumber
&\quad& + \frac{1}{\tau}\partial_{\eta}(-\sinh\eta\,T^{tz}+\cosh\eta\,T^{zz}) = 0,
}

Furthermore, $\rho_{\tau}$ and $\rho_{\eta}$ are defined as follows:
\eq{
\rho_{\tau} &=& \cosh\,\eta T^{tt}-\sinh\eta\,T^{tz} \\ 
\nonumber
&=& \delta \varepsilon \cosh\eta + \left(\varepsilon + P - \frac{4}{3\tau}\eta_s\right)\delta u^{\eta}\sinh\eta,\\
\nonumber
\rho_{\eta} &=& \cosh\,\eta T^{tz}-\sinh\eta\,T^{zz} \\
\nonumber
&=& \delta \varepsilon \sinh\eta + \left(\varepsilon + P - \frac{4}{3\tau}\eta_s\right)\delta u^{\eta}\cosh\eta.
}
Together with $\rho_B,\rho_Q,\rho_S$, they satisfy the conservation law:
\eq{
\int \tau d\eta \rho(\tau,\eta) = \text{const}.}
In the following quantities:
\eq{
\delta P_{\eta}&=&\left(\varepsilon+P-\frac{4\eta_s}{3\tau}\right)\delta u^{\eta},\\
\nonumber
\delta P_{x,y}&=&\left(\varepsilon+P-\frac{4\eta_s}{3\tau}\right)\delta u^{x,y},
}
will be called longitudinal and $x,y$ momentum respectively.
Using Green's functions of these equations, correlators can be expressed as follows:

\eq{\label{cab}
c_{AB}&=& \int d\tau_j \tau_j d\eta_j G_{A A^`}(\eta_1-\eta,\tau_1,\tau_j)G_{B B^`}(\eta_2-\eta,\tau_2,\tau_j)\left(\partial_{\tau_j}+\frac{1}{\tau_j}\right)\chi_{AB}(\tau_j),}
where $A,B$ can be any of the following: $\delta E,\delta P_{\eta},\delta P_{x,y},\delta B,\delta Q,\delta S$. The term $c_{AB}$ in Equation (\ref{cab}) does not encompass the portion of correlation that should exist in the equilibrated hydrodynamic medium right from the start. This initial correlation can take on any form but is nevertheless bound by the overall conservation of charge, as outlined in Eqs. (\ref{cons-rule}). In the subsequent discussion, we will exclude this initial correlation, and discuss it as a distinct element that can be independently studied.

\subsection{Projecting Correlations onto Final-State Hadrons}

The previous subsections describe how one might find the correlations $\langle \delta j^\mu_i(x)\delta j^\nu_k(x')\rangle$, which are functions of space-time. These correlations must then be projected onto final-state particles. Techniques for this have been applied for purely charge fluctuations \cite{Pratt:2017oyf,Pradeep:2022mkf}, but for this problem we need to extend those ideas to include correlations involving momentum and energy. As was done with the case with charges, the techniques will be based on the Cooper-Frye equation, where emission from the hyper-surface element for uncorrelated particles from a small hyper-surface element $\delta\Omega_\mu$ is given by 
\begin{eqnarray}
\delta dN_h&=& (p\cdot \delta \Omega)\frac{d^3p}{E_p}f_h(\vec{p},x),
\end{eqnarray}
where $f_h(\vec{p},x)$ is the phase space density of hadrons of type $h$ and momentum $\vec{p}$ at space-time point $x$, with $x$ being at the hyper-surface element. An efficient algorithm for generating particles consistent with a thermal phase space density, including viscous corrections, is described in \cite{Pratt:2010jt}. The approach here will be to first integrate over each source point $X$. For each point $X$, one then generates completely uncorrelated pairs of hadrons. The two hadrons are uncorrelated with the source point $X$ and are uncorrelated with one another. The correlation function in momentum space is constructed by incrementing the correlation functions by a correlation weight. 

To calculate the weight, one can consider a small amount of energy, momentum, and charge, $\delta Q_i$, that passes through a hyper-surface element. For a thermal distribution, the phase-space density is altered by modifying the Lagrange multipliers associated with each conserved quantity,
\begin{eqnarray}\label{eq:fofbeta}
f_h(\vec{p},x)&=&f_h^{(0)}(\vec{p},x)\exp(-\delta\beta_iq_i(h,\vec{p})).
\end{eqnarray}
Here, $i$ refers to the seven generalized charges of the particle of species $h$ and given energy and momentum, as described at the beginning of this section. The quantity $q_i(h,\vec{p})$ are the seven generalized charges of a single hadron. The seven components of $\delta\beta$ are determined by fixing the seven quantities, $\delta Q_i$, which are the charges passing through the hyper-surface volume $V$.
\begin{eqnarray}
\delta Q_i&=&\sum_h\int \frac{d^3p}{(2\pi)^3E_p}(p\cdot d\Omega)q_i(h,\vec{p})\delta f_h(\vec{p},x)\\
\nonumber
&=&-\sum_{h,j}\int \frac{d^3p}{(2\pi)^3E_p}(p\cdot d\Omega)q_i(h,\vec{p})q_j(h,\vec{p})f_h^{(0)}(\vec{p},x)\delta\beta_j.
\end{eqnarray}
If the frame of hyper-surface dissolution is the fluid frame, can identify $p\cdot d\Omega/E_p$ as a volume $V$, and the r.h.s. can be identified with the susceptibilities,
\begin{eqnarray}
\delta Q_i&=&-V\chi_{ij}\delta\beta_j,\\
\nonumber
\delta\beta_i&=&-\chi^{-1}_{ij}\delta Q_j/V=-\chi^{-1}_{ij}\delta\rho_j.
\end{eqnarray}
Inserting this into Eq. (\ref{eq:fofbeta}), one then has the alteration of the phase space density
\begin{eqnarray}
\delta f_h(\vec{p},x)&=&f_h^{(0)}\left[-q_i(h,\vec{p})\chi_{ij}^{-1}(x) \delta \rho_j\right].
\end{eqnarray}
Everything in the square brackets can be considered as a weight due to the additional generalized charge density $\delta\rho_j$.

For each differential space-time volume of the source, one can then generate two independent particles according to the phases-space densities at the hyper-surface, $f^{(0)}(\vec{p},x)$. One then assigns a weight to the pair,
\begin{eqnarray}
w(\vec{p}_1,x_1,\vec{p}_2,x_2)&=&\int d^4X~
q_i(h_1,\vec{p}_1)\chi_{ij}^{-1}(x_1)d\Omega_{1,\mu}
G^{\mu\mu'}_{j'j}(X,x_1) u_{\mu'}(X)\\
\nonumber
&&S_{j'k'}(X)u_{\nu'}(X) G^{\nu\nu'}_{k'm}(X,x_2)
\chi_{mn}^{-1}(x_2)d\Omega_{2,\nu}q_n(h_2,\vec{p}_2).
\end{eqnarray}
One must decide whether to perform the integral over $d^4X$ with Monte Carlo methods. For the case considered in the next section, one can exploit the symmetries, that the source function depends only on the proper time $\tau$ and that the Green's functions depend on the relative rapidities, to simplify the integral. But, if these approaches were to be applied more generally, the integral above would require; a much more nuanced approach.

One advantage of this approach is that weighted pairs are generated by Monte Carlo, so it is straight forward to incorporate the effects of decays. Hadronic decays can be simulated with the same weight from the parents being assigned to the daughter particles. All the pairs of daughters, with one daughter chosen from each parent, would be assigned the same weight as the parental pair. It is also rather straight-forward to consider a variety of binnings of the correlation. For example if one were to calculate correlations in transverse energy as a function of relative pseudo-rapidity, one would increment the bin with that relative pseudo-rapidity by the weight multiplied by the product of the two transverse energies.

\section{Results}\label{sec:results}
\subsection{Equations of State}

As an equation of state (EoS), we have used the free energy model for a hadronic mixture (details can be found in the appendix). The background charges were fixed to reproduce $Q/B=0.4$, as in heavy lead or gold nuclei, and $S=0$. All trajectories start at $\varepsilon=3~\text{GeV} \text{fm}^{-3}$ and evolve for $10~\text{fm}/c$. The final temperature falls within $150 \pm 2$ MeV. The initial baryon density was chosen at $8n_0$, where $n_0=0.16~\text{fm}^{-3}$ represents normal nuclear density, and the viscosity lies in the range of $(6-8)\frac{s}{4\pi}$, as expected for the hadronic medium. This keeps the temperature of the medium below $190$ MeV, within the range where hadronic type of the equation of state and transport can be hypothetically feasible. The diffusion coefficient $D$ was chosen to be lower than the values predicted from lattice QCD at vanishing baryon charge, aligning more with hadronic phase simulations, with $D=\frac{1}{8\pi}T$. However, in principle, the non-diagonal diffusion matrix should be applied to charges, as diffusion predicted from different kinetic models is strongly influenced by particle mass~\cite{Fotakis:2019nbq}. Electric charge is expected to be carried mainly by light mesons, and baryon charge by heavier particles. Strangeness is interesting as it includes heavy mesons and baryons, and it is thus  correlated with baryon number. Therefore, one might expect a strong sensitivity of the diffusive properties to the hadronic composition of the medium.
\begin{figure}[t!]
\includegraphics[width=.49\textwidth]{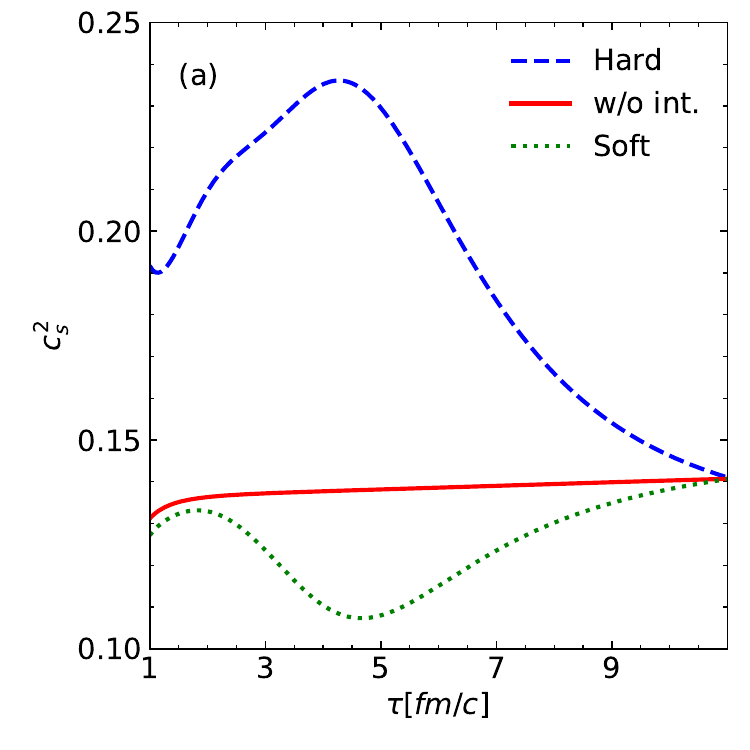}
\caption{\label{eos-plots-cs2} The speed of sound is displayed for a non-interacting (ideal) hadron resonance gas and a hadronic fluid with two illustrative cases for interactions. The equation of state that includes interactions can be either stiffer (Hard-EoS) or softer (Soft-EoS) compared to the non-interacting one (Ideal HRG).
}
\end{figure}
\begin{figure}[t!]
\includegraphics[width=.49\textwidth]{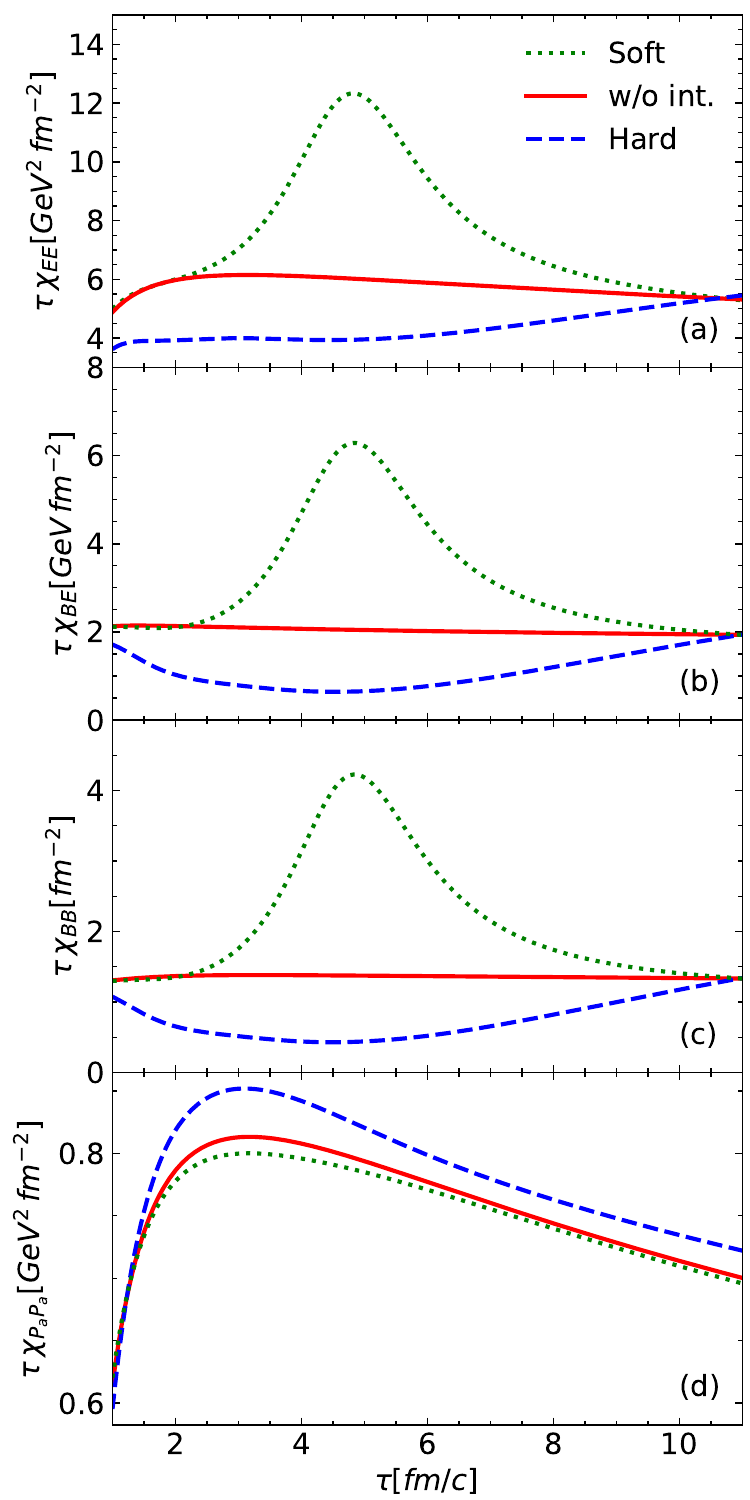}
\caption{\label{eos-plots-chi}The susceptibilities of energy-energy (a), energy-baryon density (b), baryon-baryon densities (c) and momentum-momentum (d) are shown for the three equations of state that were considered. The difference between different EoS is reflected in the susceptibilities of conserved charges. 
}
\end{figure}

Using the formalism in Appendix \ref{sec:helmholtz}, two equations of state were constructed using simple parametric forms for the density dependence of the free energy. The two interactions were not meant to be particularly realistic, but were constructed to provide the means to compare the difference between a soft and stiff equation state. With that goal in mind, the equations of state were constructed to be represent the outer limit of expectations, rather than to represent some sort of best guess. The free-energy densities were chosen to depend only on baryon density $\rho_B$, and to have the form
\begin{eqnarray}
f_{\rm int}&=&\sum_{n=1,2}A_n\left[y^{1/3}-1\right],\\
\nonumber
y&=&1+\left(\frac{\rho_B-\rho_0}{\rho_n}\right)^3.
\end{eqnarray}
Here, $\rho_0$ was set to the break up density so that the equation of state, including fluctuations, would return to those of a non-interacting hadron gas at that point. The parameters $A_1,A_2,\rho_1$ and $\rho_2$ were then adjusted to set the stiffness of the equation of state in various density ranges. For both the stiff and soft  equations of state, $\rho_1$ and $\rho_2$ were set to 0.2 and 0.5 fm$^{-3}$ respectively. For the stiff equation of state the stiffness parameters were set to $A_1=0.05,A_2=0.1$ GeV/fm$^3$, and for the soft equation of state $A_1=-0.02, A_2=0.01$ GeV/fm$^3$.  The soft equation of state has a phase transition with a critical point at $T=100$ MeV at a density near $2\rho_s$. This rather extreme example should should help bracket the range of fluctuations one might wish to consider.

The initial baryon density was chosen so that the initial baryon density was eight times nuclear saturation density, $\rho_s=0.16$ fm$^{-3}$, which because the baryon density fell as $1/\tau$ always yielded the same final baryon density, $\rho_0=(8/11)\rho_s$, regardless of the equation of state. The initial temperature, $T_i$ was adjusted for each equation of state, and for each choice of viscosity, so that the final temperature, $T_f$, was 150 MeV. By construction, the speed of sound matches that of a non-interacting hadronic mixture at and below $\rho_0$. In Fig. \ref{eos-plots-cs2} the speed of sound is displayed for the three equations of state. For the non-interacting case the speed of sound stays fairly steady, with $c_s^2\approx 0.14$. For the stiff equation of state the maximum speed of sound for the evolutions considered here was approximately, $c^2_{s,{\rm max}}\approx 0.24$, while for the soft equation of state the speed of sound dropped to $c^2_{s,{\rm min}}\approx 0.11$.

Susceptibilities of a few selected conserved charges can be observed in Fig. \ref{eos-plots-chi}. Susceptibilities are multiplied by $\tau$ as the volume of a hydrodynamic cell increases proportionally to $\tau$ in the Bjorken model. Thus, in this case, the source functions are precisely the rates at which the curves in Fig. \ref{eos-plots-chi} rise or fall. The susceptibilities involving energy or baryon number all have strong peaks or valleys for the same conditions when the speed of sound has a maximum or minimum.

xRather remarkably, the momentum-momentum susceptibility (multiplied by $\tau$), $(P+\epsilon)T\tau$, is relatively featureless and is thus rather insensitive to the equation of state. This probably owes itself to the fact that the interaction is a function of baryon density, and is of course an energy, but does not much alter the momenta of particles. It would be interesting to understand whether a different class of interaction, e.g. one that alters the degrees of freedom, might manifest itself by providing more structure for this case. Even if the flatness of momentum-momentum susceptibility results in little source function, there would still be a contribution from the initial thermalization, $\tau< 1$ fm/$c$. The correlation of the transverse components, e.g. $\mean{\delta P_x \delta P_x}$, is expected to spread diffusively \cite{Pratt:2016lol,Pratt:2010zn}, with the shear viscosity playing the role of the diffusion constant. Thus, the correlations of transverse momentum might be less sensitive to phase structure, but more sensitive to the viscosity. Equations of state featuring a critical point have an adjacent mixed phase on the phase diagram ($(\rho,T)$). For the soft equation of state shown here, phase separation can ensue, but the critical temperature is 100 MeV, and the density-temperature trajectories are well above the critical point, and therefore outside the mixed-phase region. 

Within the mixed phase, or coexistence, region, susceptibilities are undefined, i.e. fluctuations no longer scale linearly with volume ~\cite{Kuznietsov:2022pcn,PhysRevC.102.024908}. One might then anticipate the occurrence of nucleation or cavitation, representing the dynamic formation of a new phase. This process unfolds relatively slowly and challenges the application of traditional equilibrium thermodynamics methods like Maxwell constructions. In such cases, a metastable state can be contemplated, corresponding to the gradual growth of nucleation or cavitation~\cite{Kuznietsov:2023iyu,Sorensen:2020ygf}. Within the spinodal region, mechanical instability triggers a phenomenon known as spinodal decomposition, where two phases rapidly segregate. Ideally, this can be incorporated into our analysis by integrating the appropriate hydrodynamic background. Specifically, the inclusion of surface tension and energy terms in hydrodynamics can provide an accurate description of the mixed phase (for further insights, see \cite{PhysRevC.96.044903}). The approach here would then certainly not apply if the density-temperature trajectory were to pass through the coexistence region of the phase diagram. Nevertheless, it is worth noting that substantial structure in susceptibilities are expected, even if merely passes the vicinity of the phase transition region. 

In equilibrated systems correlation lengths diverge near the critical point \cite{Mukherjee_2015}. In that limit the finite lifetime of the collision prevents the system having equilibrium. The assumption here that the short-range correlations can be assigned the strength of the equilibrium susceptibility would then need to be adjusted. In principle, the current approach could be modified to account for this by treating that strength as a dynamic quantity. Non-equilibrium considerations of the local correlations  and the accompanying hydrodynamic response are intriguing, but they will not be considered in this study. 

\subsection{Green's Functions}

\begin{figure*}[t!]
\includegraphics[width=.49\textwidth]{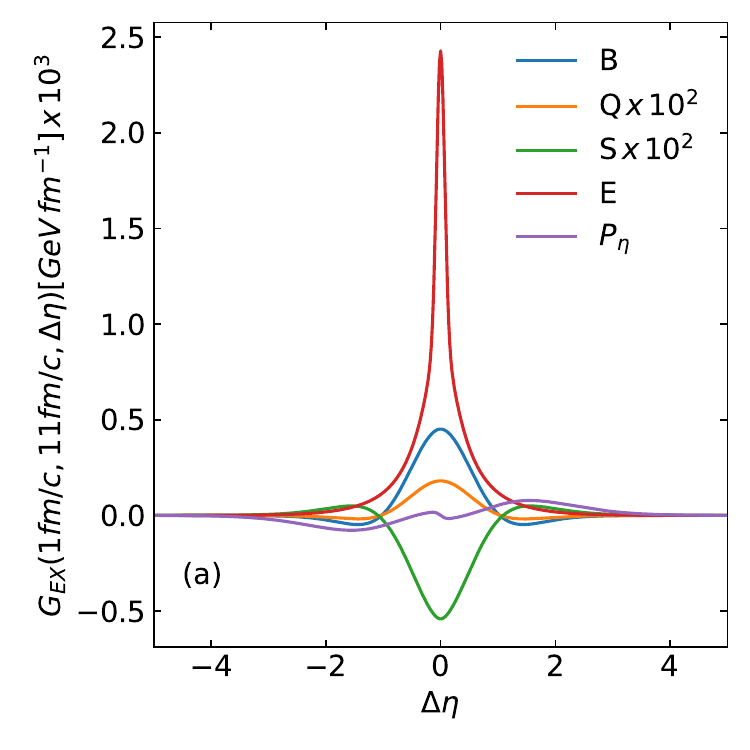}
\includegraphics[width=.49\textwidth]{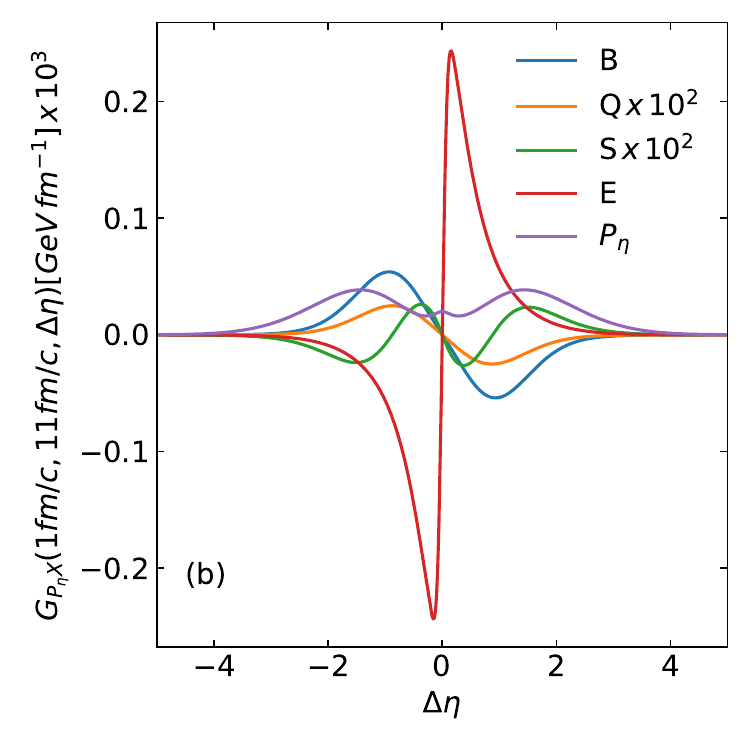}
\includegraphics[width=.49\textwidth]{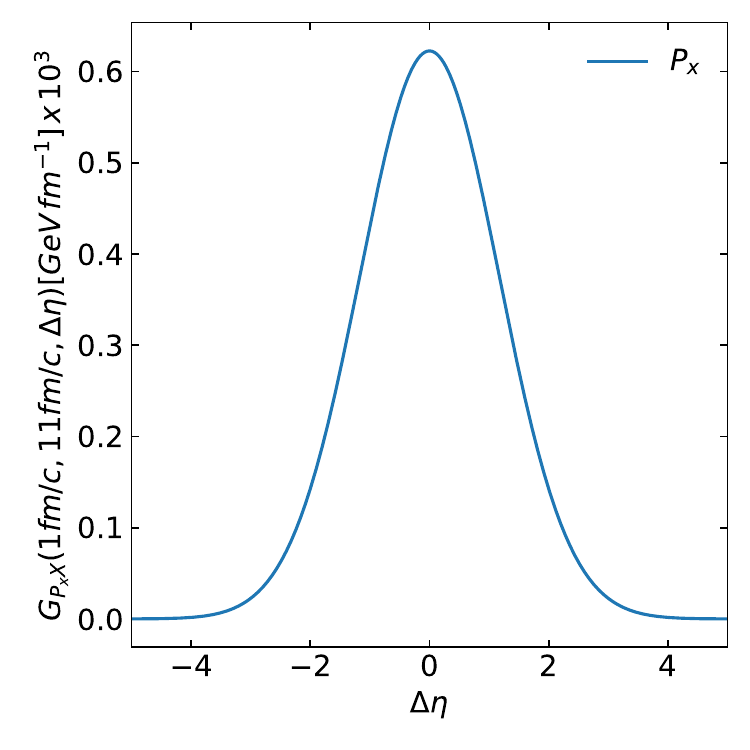}
\includegraphics[width=.49\textwidth]{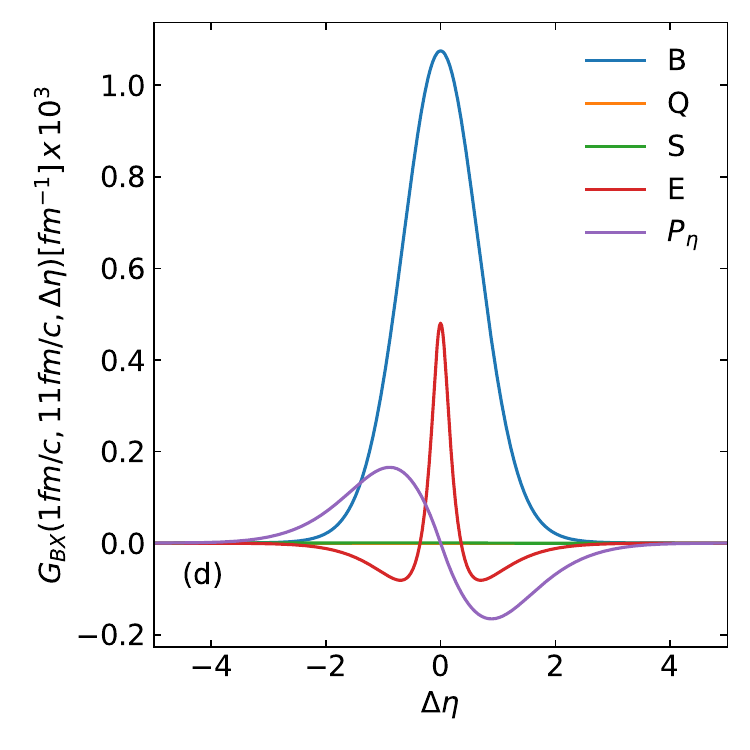}
\caption{\label{Greens-func} Green's functions, denoted as $G_{YX}(\tau_{\text{source}}=1\, \text{fm}/c, \tau=11\, \text{fm}/c, \Delta \eta)$, represent the response to a perturbation in $X$, initiated at $1\, \text{fm}/c$ at $\eta=0$, and observed in $Y$ at $\Delta \eta$. Panels $a,b,c,d$ display $G_{YX}$ for $Y=E,P_\eta,P_x$ and $B$ respectively. For each $Y$, each panel shows the non-zero Green's functions for the quantities described in the legends. 
These functions are computed for the soft version of the equation of state, and further details can be found in the accompanying text.
Cross-correlations between different conserved charges are considered, including longitudinal energy-momentum, baryon, and electric charges. To enhance visibility, some of the functions were scaled by large factors.
}
\end{figure*}

In order to obtain a solution, a set of Green's functions for point sources in $(\tau,\eta)$ must be generated. Taking advantage of the boost invariance of the Bjorken flow, we consider only $(\tau,0)$ and derive a solution for a source at any $\eta$ using Lorentz transformations. Green`s functions were computed by assigning initial condition at $\tau_{\rm start}$ in some charge $X$:
\eq{ 
G_{XX}(\tau=\tau_{\rm start},\eta)=\frac{1}{\sqrt{2\pi
\sigma^2_{\eta}}\tau}\exp\left\{-\frac{\eta^2}{2\sigma^2_{\eta}}\right\},
}
as $\sigma_{\eta}\to 0$, $G_{XX}(\tau=\tau_{\rm start},\eta)\to {\delta(\eta)\over \tau}$ approaches solution for point source (Green`s function by definition). Using very small but finite sigma regularizes our system by regularizing high momentum modes. 
Figure (\ref{Greens-func}) depict the response to a Gaussian source with $\sigma^2_{\eta}=0.01$ and $\tau_{\rm start}=1 \text{fm}/c$ at $\tau=11 \text{fm}/c$. It is noticeable that a perturbation in one charge leads to a response in another. This can be comprehended from the expressions, Eq. (\ref{eq-e}-\ref{eq-px}). Equation (\ref{eq-rho}), in its second term, describes the convection of background charge by a perturbation of velocity. This effectively couples the response in charges to the response in energy-momentum. Equations (\ref{eq-e}) and (\ref{eq-pz}) are influenced by the response to charge fluctuations through the dependence on the equation of state $P(\varepsilon,\rho_B,\rho_Q,\rho_S)$.

One expects two forms of response: two sound wave modes linear in wave number and one quadratic mode that corresponds to dissipation. Sound waves propagate at the adiabatic speed of sound of the medium and correspond to the long-range part of correlations. Parts that are due to dissipative processes in the medium are relatively localized and extend over approximately 1-2 units of space-time rapidity. This is on the order of the thermal spread of momentum rapidity for matter at freeze-out. Therefore, it is interesting to observe to what extent the signal of hydrodynamic evolution will survive particlization.

\subsection{Correlations}
\begin{figure*}[t!]
\includegraphics[width=.99\textwidth]{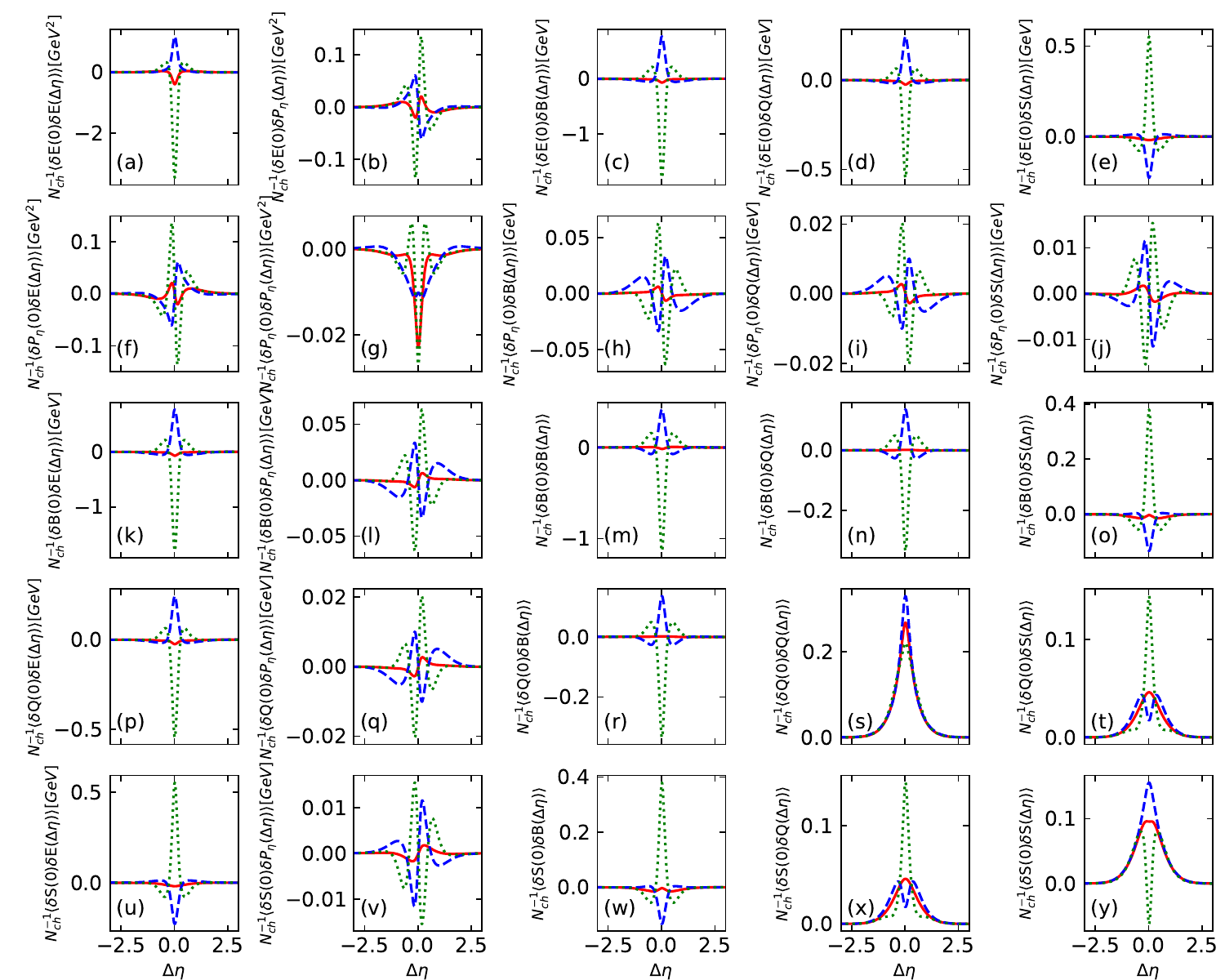}
\caption{\label{corr-init} The matrix of two-point correlations for energy, longitudinal momentum, and baryon, electric, and strange charges is presented as a function of space-time rapidity distance at the freeze-out time $\tau=11 \text{fm}/c$. The considered equations of state include the ideal hadron resonance gas (in blue), hard hadronic fluid (in orange), and soft hadronic fluid (in green). Transverse fluctuations of the system are decoupled from other observables and can be found in a separate plot.
$N_{ch}$ is charged particle multiplicity per unit of $\eta$ as predicted from statistical hadronization at $T=150\, MeV$ with transverse area $S=100\,fm^2$.}
\end{figure*}

Three equations of state are compared here: the ideal Hadron Resonance Gas (iHRG) and interacting hadron fluids, with one being harder and the other softer than the iHRG. Fluctuations involving energy, longitudinal momentum, baryon density, electric charge, and strange number densities are separated from those involving transverse momentum, as correlations between them vanish. The longitudinal hydrodynamic correlations are presented in Figure (\ref{corr-init}), and they display noticeable sensitivity to the chosen equation of state.
\begin{figure*}[t!]
\includegraphics[width=.99\textwidth]{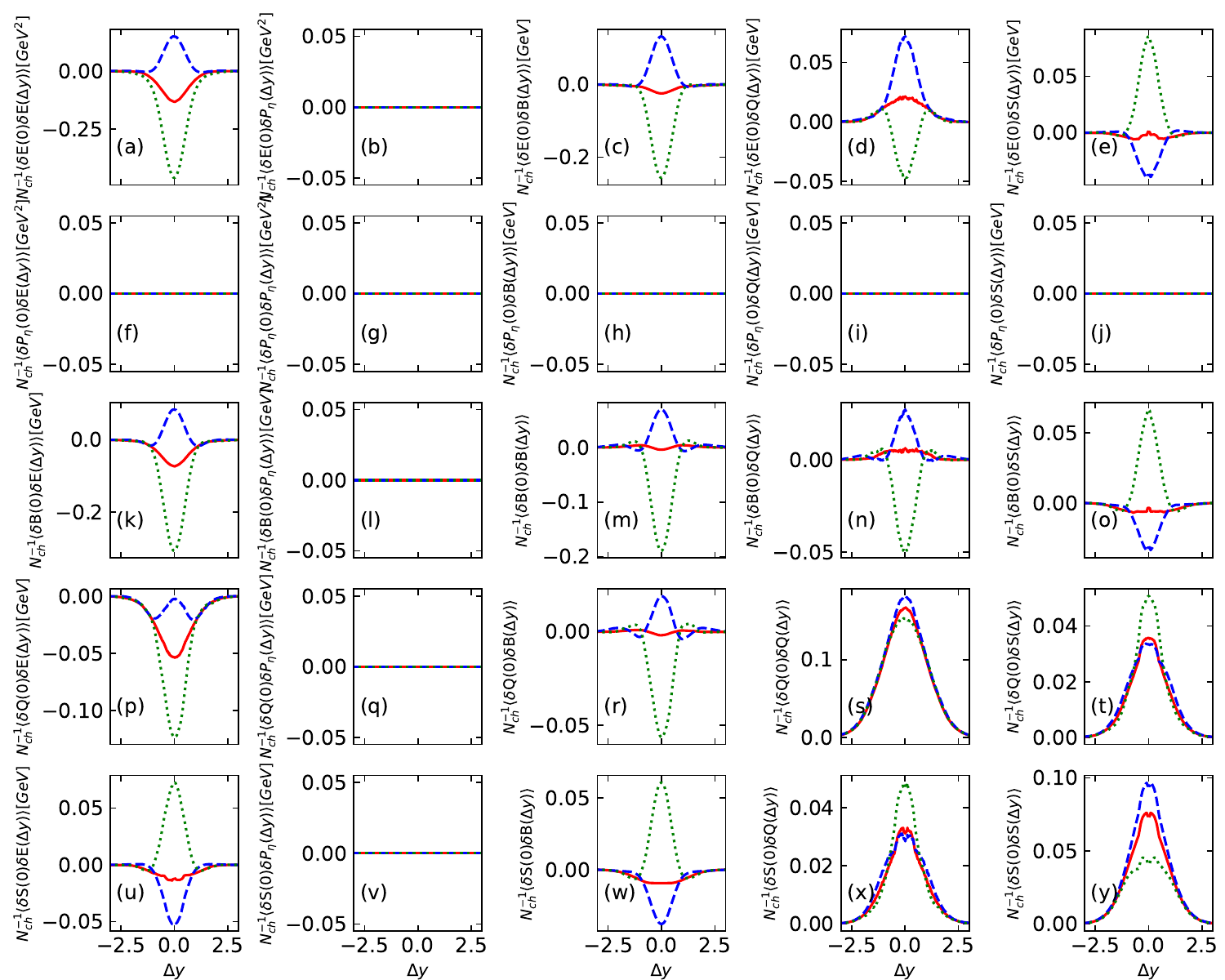}
\caption{\label{corr-out}The matrix of two-point correlations for energy, longitudinal momentum, and baryon, electric, and strange charges is presented as a function of momentum rapidity at the freeze-out time $\tau=11 \text{fm}/c$, computed after sampling all charged particles. The equations of state considered include the ideal hadron resonance gas (in blue), interacting hadronic fluid (in orange), and a fluid with a phase transition (in green).
It's important to note that transverse fluctuations of the system do not affect the longitudinal observables, and momentum fluctuations completely cancel out. $N_{ch}$ is charged particle multiplicity per unit of $\eta$ as predicted from statistical hadronization at $T=150 \,MeV$ with transverse area $S=100\,fm^2$.
}
\end{figure*}

Electric and strange charges (s),(y) are only weakly influenced by the baryon mean-field used. However, some effects are observable around $\Delta \eta \approx 0$. In this case, a harder equation of state produces a stronger peak, whereas a softer equation of state suppresses the peak in a small region. These differences disappear at the tails of the correlations. This effect is more pronounced in the strange charge compared to the electric charge. This can be attributed to the transition from a medium that includes a large number of strange and charged mesons at high temperatures to a medium dominated by cold baryons at freeze-out. Baryons, being more sensitive to the equation of state, contribute to the central part of the correlation function, while the tails are created much earlier in the collision. The cross-correlation between electric charge (Q) and strange charge (S) exhibits a similar behavior (t) and (s), but here the peak structure changes from a single maximum for an ideal gas to two peaks for the Hard-EoS or a much stronger peak for the Soft-EoS.

The baryon charge (m) exhibits more pronounced features. In particular, the value at the peak, $\Delta \eta = 0$, changes from negative to positive when transitioning from the iHRG to the Soft-EoS to the Hard-EoS. Interactions also amplify both the maximal and minimal values of the correlation function. This can be understood from Figure (\ref{eos-plots-chi}), where the rates of change of $\tau \chi(\tau)$ are much higher for the equation of state with interactions and sometimes have a different sign compared to the iHRG. The spread of the correlation here is around one unit in space-time rapidity. Correlations of the baryon charge with electric or strangeness experience similar effects from interactions (r),(w),(n),(o).

The longitudinal momentum correlation $\langle \delta P_\eta \delta P_\eta\rangle$ (g) for the iHRG is much narrower compared to the interacting version. It is evident that, in this case, the correlation spreads further for a hard equation of state (higher speed of sound) and less for the soft equation of state (smaller speed of sound). The momentum correlation with other charges appears to be proportional to the rate at which susceptibility changes.

Energy correlations (a) have a width of about half a unit of rapidity. The value at the peak changes sign for the hard and ideal hadronic medium, for the $\mean{\delta E\delta E}$ correlation. It's worth noting that the transport coefficients corresponding to viscosity for different equations of state are almost identical. It seems that the sound wave part of energy gets suppressed and transferred into momentum. It is interesting to see how this transformation manifests in the correlations at particlization.

Balancing correlation of the transverse momentum, Figure (\ref{ptpt}) (a), is weakly influenced by the equation of state. It is noteworthy that in this case, the correlations spread quite far, covering nearly 5 units of rapidity. The Hard-EoS has the largest width, while the Soft-EoS has the smallest.
\subsection{Freeze-out}

\begin{figure*}[t!]
\includegraphics[width=.49\textwidth]{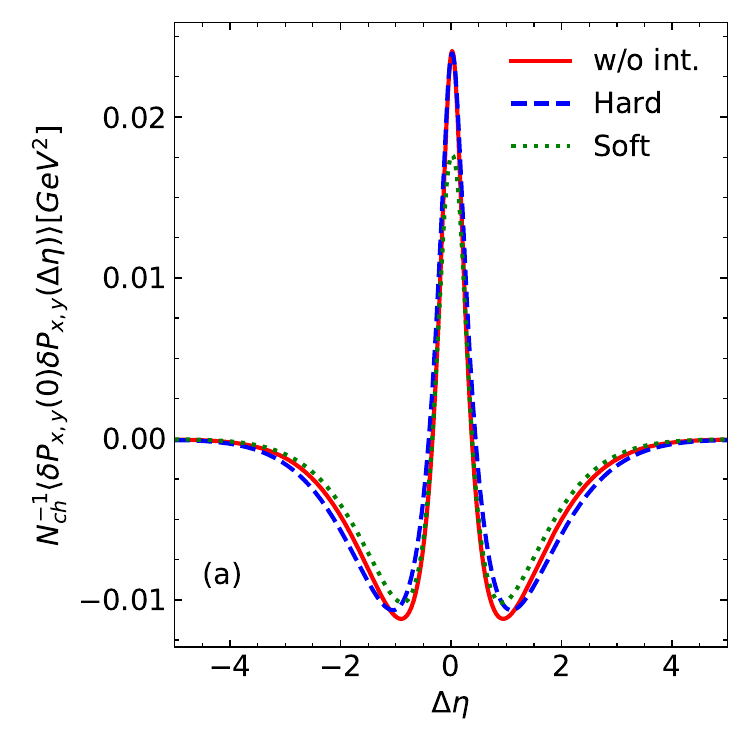}
\includegraphics[width=.49\textwidth]{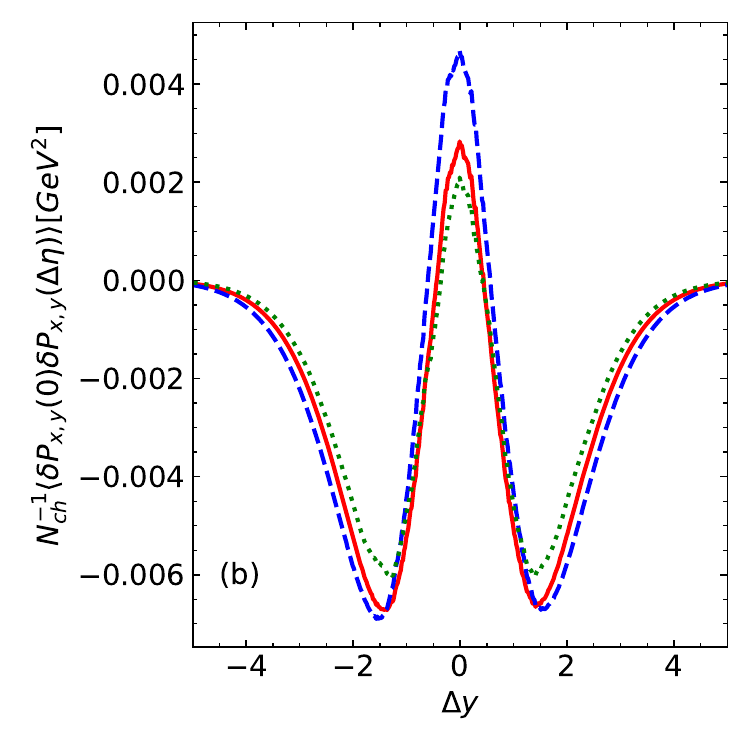}
\caption{\label{ptpt} Balancing part for the correlations of transverse momentum $P_{x,y}$ as function of space time (left) or momentum (right) rapidity differences. As predicted from hydrodynamic evolution with thermal sources before (left) and after particle creation (right). $N_{ch}$ is charged particle multiplicity per unit of $\eta$ as predicted from statistical hadronization at $T=150\, MeV$ with transverse area $S=100\,fm^2$.}
\end{figure*}
The final form of the balancing part, computed for all charged particles at freeze-out, is depicted in Figure (\ref{corr-out}) and Figure (\ref{ptpt}) for $E$, $P_{\eta}$, $B$, $Q$, $S$, $P_{x}$, and $P_{y}$, respectively.

Before particlization, the $P_{x}$ and $P_{y}$ correlations exhibit a strong positive peak at mid-rapidity with negative dips around it, Figure (\ref{ptpt}) (a). However, after particlization, the value at the peak is suppressed, while the dips are enhanced, Figure (\ref{ptpt}) (b). The differences between the equations of state become more pronounced, especially in the case of the Hard-EoS, where the width is much larger than in the other two cases.

In $(\eta, \phi)$ coordinates, correlations in transverse momentum have undergone experimental investigation, serving as a sensitive probe for viscosity. One notable feature that has been observed is the ``Ridge'' in rapidity. Correlations in transverse momenta are influenced by various physical phenomena, in addition to momentum conservation. These include resonance and cluster decays, radial flow effects, anisotropic flow effects, initial state fluctuations, and modified jet fragmentation  \cite{Bozek:2012is,Bozek:2012en}. All of these effects, with the exception of jets, can be incorporated into the (3D+1) hydrodynamic modeling of the system. Jets are formed early in the collision and are not typically part of the collective motion. However, jets interact with the bulk of the QCD medium, serving as sources of additional correlation. Therefore, it is important to understand the extent to which jets contribute to the $p_t$ correlation. As evidenced by the comparison of our results with data, the hydrodynamic history and balancing correlation should account for a significant portion of the total correlation. The investigation of bulk-jet interaction is indeed an intriguing and important topic.

While the other correlations are not as wide as the transverse momentum correlation, the width of the balancing parts increases to 2 units of rapidity. This is primarily attributed to the thermal spread of particles in momentum rapidity, denoted as $y=\eta+\delta y$, with $\delta y$ having values concentrated around the range of $[-0.5, 0.5]$. This smearing effect blurs the features and results in much smoother correlation. 

The balancing parts for strange and electric charges mostly preserve their features, differing at mid-rapidity but converging towards the tails. They exhibit a simple unimodal structure where a stronger hydrodynamic response leads to the generation of more balancing correlation. Since we have only considered balancing correlation created during the hydrodynamic stage, the normalization appears to be proportional to the difference in susceptibilities at $\tau=1$ and $\tau=11 \text{fm}/c$, which, as can be inferred from Figure (\ref{eos-plots-chi}), is larger for the Hard-EoS. This difference is notably more significant for the baryon charge. The minima surrounding the peak in this case become less pronounced when compared to the correlation before freeze-out. In the case of the ideal Hadron Resonance Gas (iHRG), the balancing part seems to remain almost constant around 0, completely losing any rapidity dependence that was present before.

When longitudinal momentum becomes involved, the correlation vanishes. This is a consequence translation invariance in $\eta$. If other than periodic boundary conditions were used that would not be true anymore. If one were to consider a system that is not exactly transnational invariant, Green's functions would become functions of both rapidities (source and response), and not just their difference. 

For energy, the width is now comparable to that of electric or strange charges, and it is even larger than the width of the baryon charge. Additionally, it appears that a hard equation of state becomes much more spread compared to the other equations of state considered. This phenomenon reflects the presence of hydrodynamic modes in response to the initial perturbation.

\section{Conclusions}\label{sec:conclusions}

Here, the entire $7\times 7$ matrix of correlations or conserved quantities was considered as a function of the measure of separation along the beam axis. For correlations in coordinate space, correlations were binned as a function of relative spatial rapidity, $\Delta\eta$, whereas for the final-state correlations, the momentum rapidity difference $\Delta y$ was chosen. An extremely simple model was used here. Boost invariance was assumed, which is a poor choice at lower beam energies, and transverse flow was ignored, which is a poor assumption at any energy. The equations of state applied were also not particularly physical. The correlations are seeded by the evolution of the susceptibilities and evolve according to viscous hydrodynamics and diffusion. At a fixed breakup time, correlations were projected onto those of free-streaming hadrons and their decay products, ignoring any additional contributions from the hadron phase aside from those from decays. For example, baryon annihilation is ignored here. Only thermal correlations were considered, which ignore those from the initial state. Initial state correlations can be driven by jets or minijets, or any local energy-density fluctuations due to the stopping mechanism. Though the integrated strength of such correlations is also constrained by conservation laws, their initial structure is not well understood and might well be much stronger than the thermally driven correlations discussed here.

Nonetheless, these calculations were sufficient to illustrate the potential of such correlations to constrain the equation of state and to answer some basic questions. Given that overcoming the shortcomings enumerated in the previous paragraphs requires a substantial commitment from both the theory and experimental communities, this study should, hopefully, clarify whether such a significant investment might be worthwhile. Here, we summarize the insights gained and the lessons learned from this study.
\begin{enumerate}
	\item  It is possible to consistently model the entire matrix of correlations. This requires simultaneous consideration of all seven conserved quantities. Because boost invariance was assumed, and transverse momentum was ignored, the calculations here were greatly simplified, but the paradigm presented here is certainly extendable to more realistic cases. More realistic calculations would first require careful consideration of the algorithms to handle the additional dimensionality, and adding transverse flow would then allow the correlations involving $P_x$ and $P_y$ to mix with the other five quantities.

	\item All the correlations were significantly sensitive to the equation of state and transport coefficients, and such correlations survived the particlization of being projected from the hydrodynamic stage to free-streaming, but decaying, hadrons. Correlations often had non-trivial structures, and the widths in rapidity were $\lesssim 1$ unit of rapidity. Correlations from the initial state, which were ignored here, should have structures characterized by $\gtrsim 1$ unit of rapidity, because of the greater time they have to spread.
 
The strength of thermal correlations for central collisions was found to be significant, similar in magnitude to those seen experimentally in $p_t-p_t$ correlations \cite{PhysRevC.99.044918}, or correlations often referred to as the ``Ridge'' \cite{PhysRevC.80.064912}. Thus, the thermally sourced correlations considered here are not overwhelmed by those of the initial state. Given their narrower structure, there is hope they might somewhat separable from those of the initial state, but indications are that this would require significant modeling and analysis, both experimentally and theoretically.

	\item Such correlations are highly intertwined, especially at lower beam energy where the net baryon density significantly differs from zero. At the highest energies, charge-charge correlations do not mix with those involving energy and momentum. For collisions studied at the SPS at CERN, or for the Beam-Energy-Scan at RHIC, correlations appear between charges, such as baryon number, and energy. Energy and momentum correlations then mix from hydrodynamics. Thus, it is crucial to simultaneously consider the entire matrix of correlations, which would include such quantities as the correlation between baryon number and transverse energy.

\item Calculating correlations given the equation of state, transport coefficients, and initial state effects already represents a significant challenge. Performing the inverse calculation, where one constrains or infers the equation of state from measurements, is even more daunting. Given the current uncertainties in the initial state, especially for lower beam energies, pursuing this class of correlations at the current time is probably premature for lower beam energies. However, once all the modeling elements are in place and better understood, these correlations have the potential to provide tremendous insight into the evolution of bulk properties of matter throughout a heavy-ion collision. These correlations should have especially penetrating insight into any phase transitions.

\end{enumerate}

	\begin{acknowledgments}
Supported by the Department of Energy Office of Science through Grant No. DE-FG02-03ER41259. 
	\end{acknowledgments}

\appendix

\section{Sum Rules}

The correlation functions in this study, along with the Green's functions are all constrained by energy, momentum and charge conservation. These constraints were rather easily expressed in Euclidean space, but in Bjorken coordinates they are more subtle. Energy and momentum transform into one another, because Bjorken coordinates involve Lorentz boosts, so the simple sum rules, such as the correlations integrating to the susceptibilities, need to be restated. This is the goal of this appendix.

In Bjorken coordinates the proper time $\tau$ and the spatial rapidity, $\eta$ replace the Euclidean coordinates of time $t$ and the longitudinal position $z$ through the transformation,
\begin{eqnarray}
\tau&\equiv&\sqrt{t^2-z^2},\\
\nonumber
\eta&\equiv&\tanh^{-1}(z/t).
\end{eqnarray}
The collective velocity is $v_z=\tanh\eta$. For boost invariant systems, all local thermodynamic quantities depend only on $\tau$, including the criteria for decoupling. Thus, it makes sense to focus on correlations at fixed $\tau$. At fixed $\tau$ correlations, $\langle\delta A(\eta_1)\delta B(\eta_2)\rangle$ depend only on the relative rapidity, assuming that $A$ or $B$ refer to the energy, momentum and charge densities expressed in the local Bjorken frame. The local Bjorken frame involves boosting with velocity $\tanh\eta$ from the laboratory frame. In the absence of fluctuations this frame is the same as the local fluid frame, though the fluid frame will differ slightly once the perturbations are considered. The goal for this section is to express the sum rules for the correlations, defined in the local rest frames, in terms of integrals over relative spatial rapidity, $\Delta\eta$. 

\subsection{Constraints for Correlation Functions}

First, we consider correlations involving the energy and the longitudinal momentum densities. For any operator $A(\eta=0)$, 
\begin{eqnarray}
\int d\eta\langle \delta \rho_E(\eta)A(0)\rangle&=&0,\\
\nonumber
\int d\eta\langle \delta M_z(\eta) A(0)\rangle&=&0
\end{eqnarray}
due to energy and momentum conservation, and $\delta \rho_E$ and $\delta M_z$ are the changes to the energy and momentum per some slice of spatial rapidity and the energy and momenta are defined in the laboratory frame. One can write $\delta \rho_E$ and $\delta M_z$ in terms of $\delta\epsilon$ and $\delta \tilde{M}_z$, which are the energy and momentum densities in the local Bjorken frame,
\begin{eqnarray}
\delta \rho_E&=&\delta\epsilon\cosh\eta+\delta\tilde{M}_z\sinh\eta,\\ \nonumber
\delta M_z&=&\delta\tilde{M}_z\cosh\eta+\delta\epsilon\sinh\eta.
\end{eqnarray}
One can define a four-vector, $\delta M^\mu=T^{\mu\nu}d\Omega_\mu$, where $d\Omega$ is a hyper-volume element. In the local Bjorken frame, for a cross-sectional area $A$, $d\Omega_\mu=A(\tau d\eta,0,0,0)$, and in the laboratory frame $d\Omega_\mu=A(\tau d\eta\cosh\eta,0,0,\tau d\eta\sinh\eta)$. Given that the break hyper-surface element, $d\Omega_\mu$, is defined by constant $\tau$, the differential four-momentum in the Bjorken frame is $\delta \tilde{M}^\mu=(\delta\epsilon,0,0,\delta\tilde{M}_z)$. 

Energy and momentum conservation thus give the following constraints:
\begin{eqnarray}\label{eq:EPcons}
\int d\eta \langle(\delta\epsilon(\eta)\cosh\eta+\delta\tilde{M}_z(\eta)\sinh\eta)\delta\epsilon(0)\rangle&=&0,\\ \nonumber
\int d\eta \langle(\delta\epsilon(\eta)\cosh\eta+\delta\tilde{M}_z(\eta)\sinh\eta)\delta\tilde{M}_z(0)\rangle&=&0,\\ \nonumber
\int d\eta \langle(\delta\tilde{M}_z(\eta)\cosh\eta+\delta\epsilon(\eta)\sinh\eta)\delta\epsilon(0)\rangle&=&0,\\ \nonumber
\int d\eta \langle(\delta\tilde{M}_z\cosh\eta+\delta\epsilon(\eta)\sinh\eta)\delta\tilde{M}_z(0)\rangle&=&0.
\end{eqnarray}
Next we define correlations, $C_{ab}(\eta)$ in such a way that they ignore zero range correlations, and assume the zero range correlations (particles with themselves mostly) are defined by the susceptibilities. 
\begin{eqnarray}
\langle(\delta\epsilon(\eta)\delta\epsilon(0)\rangle&=&C_{EE}(\eta)+\chi_{EE}\delta(\eta),\\ \nonumber
\langle(\delta\epsilon(\eta)\delta\tilde{M}_z(0)\rangle&=&C_{EP}(\eta),\\ \nonumber
\langle(\delta\tilde{M}_z(\eta)\delta\epsilon(0)\rangle&=&C_{PE}(\eta),\\ \nonumber
\langle(\delta\tilde{M}_z(\eta)\delta\tilde{M}_z(0)\rangle&=&C_{PP}(\eta)+\chi_{PP}\delta(\eta).
\end{eqnarray}
The susceptibility $C_{EE}$ is the specific heat (multiplied by $T^2$) and $C_{PP}=(P+\epsilon)T$. From inserting these definitions into Eq.(s) (\ref{eq:EPcons}),
\begin{eqnarray}\label{eq:C2constraints}
\int d\eta \left[C_{EE}(\eta)\cosh(\eta)+C_{PE}(\eta)\sinh(\eta)\right]&=&-A\tau\chi_{EE},\\ \nonumber
\int d\eta \left[C_{PP}(\eta)\cosh(\eta)+C_{EP}(\eta)\sinh(\eta)\right]&=&-A\tau \chi_{PP}.
\end{eqnarray}
The functions $C_{EE}$ and $C_{PP}$ are even functions in $\eta$ and $C_{PE}$ and $C_{EP}$ are odd functions, with $C_{PE}(\eta)=-C_{EP}(\eta)$. These correlations were defined in terms of operators at positions $\eta$ and zero, but all the correlations $C_{ab}(\eta_1,\eta_2)$ are functions of $\eta_1-\eta_2$ only, thus the $\eta$ dependence above can be replaced with the relative spatial rapidity, $\Delta\eta$.

Subtracting the two expressions in Eq. (\ref{eq:C2constraints}) from one another provides a particularly useful expression as it ultimately leads to expressions for the transverse energy,
\begin{eqnarray}
\label{eq:CTTconstraint}
\int d\eta\left[C_{EE}(\eta)\cosh\eta+C_{PE}(\eta)\sinh\eta
-C_{EP}(\eta)\sinh\eta-C_{PP}(\eta)\cosh\eta\right]&=&-\chi_{EE}+\chi_{PP}.
\end{eqnarray}
At large times, when the particles are free streaming, the rapidity and spatial rapidity become equal. The right-hand side of Eq. (\ref{eq:CTTconstraint}), which represents the local correlation of a particle with itself simply becomes the density of tracks per rapidity multiplied by the average $E_t^2$ per track. The left-hand correlations involving the $\delta\tilde{M}_z$ all vanish because, asymptotically, the particles have no longitudinal momentum as the rapidity $y$ approaches $\eta$. The only correlation remaining is thus the one involving the transverse energies, and 
\begin{eqnarray}
\int dy~\langle \delta \rho_t(y)\delta \rho_t(0)\rangle\cosh\eta&=&-\frac{dN}{dy}\langle E_t^2\rangle.
\end{eqnarray}
Here $\rho_t(y)$ is the transverse energy per rapidity, and the averaging, $\langle...\rangle$, ignores the correlation of a particle with itself.

One might ask how the other combination, $C_{EE}+C_{PP}$, might constrain the final state. Summing the two expressions,
\begin{eqnarray}
\int d\eta\left[C_{EE}(\eta)+C_{PP}(\eta)\right]\cosh\eta&=&(-\chi_{EE}-\chi_{PP})A\tau.
\end{eqnarray}
However, once the particles rapidity approach their spatial rapidities, one ends up with the same constraint. This is not to say there exists only one useful constraint. Both constraints provide stringent tests of a correlation calculation.

Constraints for correlations involving the transverse momentum components, or the baryon, electric and strangeness charge densities are easier to derive because the boost from the Euclidean coordinates to Bjorken coordinates does not affect $P_x,P_y$ or the charge densities. The constraints are then
\begin{eqnarray}
 \int d\eta \langle\delta\tilde{M}_x(\eta)\delta\tilde{M_x}(0)\rangle&=&\int d\eta \delta\tilde{M}_y\delta\tilde{M_y}\\
 &=&-A\tau(P+\epsilon)T^2,\\
 \nonumber
 \int d\eta \langle\delta\rho_i(\eta)\delta\rho_j(0)\rangle&=&-A\tau\chi_{ij},\\
 \nonumber
 \int d\eta \langle\delta\rho_i(\eta)\delta\epsilon(0)&=&0.
 \end{eqnarray}
 Due to odd reflection symmetries,
 \begin{eqnarray}
 \int d\eta \langle\delta M_z(\eta)\delta A(0)\rangle&=&0
 \end{eqnarray}
 for any $A=\epsilon,\rho_i,M_x$ or $M_y$. Finally, because transverse expansion is neglected, there are no cross terms involving $M_x$ or $M_y$, and
 \begin{eqnarray}
 \int d\eta \langle\delta M_x(\eta)\delta A(0)\rangle=\int d\eta \langle\delta M_y(\eta)\delta A(0)\rangle=0
 \end{eqnarray}
 for any $A$ unless $A=M_x$ or $M_y$ for the two expectations above.

\subsection{Constraints for Green's Functions due to Conservation Laws}

If one has energy density (energy per spatial rapidity), $\delta\epsilon$, at time $\tau_0$, the resulting energy, momentum and charge conservation can be expressed in terms of Green's functions,
\begin{eqnarray}
\delta\epsilon(\tau,\eta)&=&\int d\eta_0 G_{EE}(\tau,\tau_0,\Delta\eta=\eta-\eta_0)\delta\epsilon(\tau_0,\eta_0),\\ \nonumber
\nonumber
\delta \tilde{M}_z(\tau,\eta)&=&\int d\eta_0 G_{PE}(\tau,\tau_0,\Delta\eta=\eta-\eta_0)\delta\epsilon(\tau_0,\eta_0),\\
\nonumber
\delta \tilde{\rho}(\tau,\eta)&=&\int d\eta_0 G_{QE}(\tau,\tau_0,\Delta\eta=\eta-\eta_0)\delta\epsilon(\tau_0,\eta_0).
\end{eqnarray}
Considering the case where $\delta\epsilon(\tau_0,\eta_0)$ is a delta function at $\eta_0=0$,
\begin{eqnarray}
\delta\epsilon(\tau_0,\epsilon)=\delta E_{\rm tot}\delta(\eta_0),
\end{eqnarray}
the total energy is $E_{\rm tot}$, which along with momentum and charge must then be conserved. The energy, momentum and charge in the lab frame are defined as
\begin{eqnarray}
E_{\rm tot}&=&\int d\eta\left(\delta\epsilon(\eta,\tau)\cosh\eta+\delta\tilde{M}_z(\eta,\tau)\sinh\eta\right),\\ \nonumber
P_{z{\rm tot}}&=&0=\int d\eta\left(\delta\tilde{M}_z(\eta,\tau)\cosh\eta+\delta\epsilon(\eta,\tau)\sinh\eta\right),\\ \nonumber
Q_{\rm tot}&=&0=\int d\eta\delta\rho_E(\eta,\tau).
\end{eqnarray}
The momentum and energy expressions above can be derived by starting with writing the energy and momentum in terms of the stress-energy tensor,
\begin{eqnarray}
\delta M^\mu=\int ~\delta T_{\mu\nu}d\Omega_\nu,
\end{eqnarray}
where $d\Omega_\nu$ is the differential hyper volume, $\sim \tau d\eta$ in the local Bjorken frame. In that frame $\delta T_{00}=\delta\epsilon/\tau$ and $T_{0z}=\delta \tilde{M}_z/\tau$. Boosting the resulting energy-momentum four vector to the lab frame adds the factors, $\cosh\eta$ and $\sinh\eta$. 

Energy, momentum and charge conservation the reguire the following constraints on the Green's functions,
\begin{eqnarray}\label{eq:sumruleGE}
\int d\Delta\eta\left[G_{EE}(\tau,\tau_0,\Delta\eta)\cosh\eta+G_{PE}(\tau,\tau_0,\Delta\eta)\sinh\eta\right]&=&1,\\ \nonumber
\nonumber
\int d\Delta\eta\left[G_{PE}(\tau,\tau_0,\Delta\eta)\cosh\eta+G_{EE}(\tau,\tau_0,\Delta\eta)\sinh\eta\right]&=&0,\\ \nonumber
\nonumber
\int d\Delta\eta~G_{QE}(\tau,\tau_0,\Delta\eta)&=&0.
\end{eqnarray}

Similarly, one can consider initial fluctuations in the momentum density. The resulting constraints are:
\begin{eqnarray}\label{eq:sumruleGP}
\int d\Delta\eta\left[G_{PP}(\tau,\tau_0,\Delta\eta)\cosh\eta+G_{EP}(\tau,\tau_0,\Delta\eta)\sinh\eta\right]&=&1,\\ \nonumber
\nonumber
\int d\Delta\eta\left[G_{EP}(\tau,\tau_0,\Delta\eta)\cosh\eta+G_{PP}(\tau,\tau_0,\Delta\eta)\sinh\eta\right]&=&0,\\
\nonumber
\int d\Delta\eta~G_{QP}(\tau,\tau_0,\Delta\eta)&=&0.
\end{eqnarray}
For the charge,
\begin{eqnarray}\label{eq:sumruleGQ}
\int d\Delta\eta~G_{QQ}(\tau,\tau_0,\Delta\eta)&=&1,\\ \nonumber
\nonumber
\int d\Delta\eta\left[G_{EQ}(\tau,\tau_0,\Delta\eta)\cosh\eta+G_{PQ}(\tau,\tau_0,\Delta\eta)\sinh\eta\right]&=&0,\\
\nonumber
\int d\Delta\eta\left[G_{PQ}(\tau,\tau_0,\Delta\eta)\cosh\eta+G_{EQ}(\tau,\tau_0,\Delta\eta)\sinh\eta\right]&=&0.
\end{eqnarray}
Eq.s (\ref{eq:sumruleGE}), (\ref{eq:sumruleGP}) and (\ref{eq:sumruleGQ}) represent the constraints on Green's functions due to conservation laws.

Finally, the transverse momentum components are also conserved. Given that, in the Bjorken limit with no transverse expansion, they do not mix with the other components, and that they are invariant to longitudinal boosts, the constraints are simple,
\begin{eqnarray}
\int d\Delta\eta~G_{P_xP_x}(\tau,\tau_0,\Delta\eta)=\int d\Delta\eta~G_{P_yP_y}(\tau,\tau_0,\Delta\eta)&=&1.
\end{eqnarray}

\section{Incorporating Interactions into the Equation of State and Fluctuations}\label{sec:helmholtz}

The equations of state in this study use the hadron gas as a base, then add contributions to interactions. For a given equation of state the calculations require knowledge of all fluctuations. For hydrodynamic calculations the input variables are the energy density and the three charge densities, i.e., thermodynamics in the microcanonical ensemble. However, the pressure, and especially the fluctuations are most clearly defined in the grand canonical ensemble. Further, the interaction terms used here are functions of the density and temperature, i.e. the canonical ensemble.

\subsection{Fluctuations in the Grand Canonical Ensemble}

For a free gas, one can relatively easily program the pressure and fluctuations in terms of the temperature and chemical potentials. The grand canonical partition function is 
\begin{eqnarray}
Z_{GC}&=&\frac{PV}{T}=\sum_h (2S_h+1)\frac{V}{(2\pi)^3}\int d^3p~e^{-(E_p-\mu_h)/T}\\
\nonumber
&=&\sum_h (2S_h+1)m_h^2Te^{\mu_h/T}K_2(m_h/T),
\end{eqnarray}
where $m_h,S_h$ and $\mu_h$ are the mass, spin and chemical potential for a hadron species $h$. The energy density and fluctuations can then readily be obtained by taking derivatives of the grand-canonical partition function. If one wants such quantities in terms of the energy and charge densities, one can first find the temperature and chemical potentials from the energy and charge densities through a Newton's method. Once one knows the chemical potentials the fluctuations can be found quickly.

Susceptibilities and fluctuations are more conveniently defined in terms of $T$ and $\mu/T$, which is defined here as $\tilde{\mu}_a=\mu_a/T$.  First, the charge fluctuations are
\begin{eqnarray}
\chi_{ab}&\equiv&\frac{1}{V}\langle\delta Q_a\delta Q_b\rangle\\ \nonumber
&=&\frac{\partial\rho_a}{\partial\tilde{\mu}_b}.
\end{eqnarray}
Energy-energy and energy-charge fluctuations are
\begin{eqnarray}
\chi_{EQ_a}&\equiv&\frac{1}{V}\langle\delta E\delta Q_a\rangle\\ \nonumber
&=&\frac{\partial\epsilon}{\partial\tilde{\mu}_a}=T^2\frac{\partial\rho_a}{\partial T},\\ \nonumber
\chi_{EE}&\equiv&\frac{1}{V}\langle\delta E\delta E\rangle\\ \nonumber
&=&T^2\frac{\partial\epsilon}{\partial T}.
\end{eqnarray}

\subsection{Fluctuations in the Canonical Ensemble}

Adding interactions complicates the procedure. Interactions are most easily expressed in terms of density, which suggests using the canonical partition function. The Helmholtz free energy density, which is simply related to the partition function is convenient when adding interaction terms as functions of the charge density $\rho$ and the temperature $T$. However, expressing the fluctuations requires some care.

In terms of the canonical partition function the Helmholtz free energy density is
\begin{eqnarray}
f(\rho,T)&=&\frac{1}{V}(E-TS)=-\frac{T}{V}\ln Z_C.
\end{eqnarray}

The differential free energy density is, after using the identity $\delta V/V=-\delta\rho/\rho$,
\begin{eqnarray}
\delta f&=&\frac{\delta\rho}{\rho}f+\frac{1}{V}(-S\delta T-P\delta V)\\ \nonumber
&=&\frac{\delta\rho}{\rho}f-s\delta T+\frac{P\delta\rho}{\rho}\\ \nonumber
&=&\mu \delta\rho-s\delta T.
\end{eqnarray}

The last relation used the identity,
\begin{eqnarray}\label{eq:sept}
s=\beta\epsilon+P/T-\mu\rho/T.
\end{eqnarray}
Reading off the last expression,
\begin{eqnarray}
s&=&-\frac{\partial f}{\partial T},~~~\mu=\frac{\partial f}{\partial\rho},
\end{eqnarray}
where $f$ is assumed to be a function of $\rho$ and $T$. Using the definition of $f$,
\begin{eqnarray}
\epsilon&=&f+Ts=f+T\frac{\partial f}{\partial T}=-T^2\frac{\partial}{\partial T}\left(\frac{f}{T}\right).
\end{eqnarray}
Using Eq. (\ref{eq:sept}) one can write the pressure,
\begin{eqnarray}
P&=&-f+\mu\rho.
\end{eqnarray}
Summarizing, to this point we have the basic thermodynamic quantities in terms of $f(T,\rho)$:
\begin{eqnarray}
\epsilon&=&-T^2\frac{\partial}{\partial T}\left(\frac{f}{T}\right),\\ \nonumber
s&=&-\frac{\partial f}{\partial T},\\ \nonumber
\mu&=&\frac{\partial f}{\partial \rho},\\ \nonumber
P&=&-f+\mu\rho.
\end{eqnarray}

\subsubsection{Charge Susceptibilities in the Canonical Ensemble}

To see how to express such quantities in the canonical ensemble, one may consider the matrix
\begin{eqnarray}
M_{ab}&\equiv\frac{\partial\tilde{\mu}_a}{\partial\rho_b}.
\end{eqnarray}
The matrix is simply the inverse of $\chi_{ab}$,
\begin{eqnarray}
\delta\rho_a&=&\chi_{ab}\delta\tilde{\mu}_b,\\ \nonumber
\delta\tilde{\mu}_a&=&M_{ab}\delta\rho_b.
\end{eqnarray}
If the lower expression is multiplied by the inverse of $M$,
\begin{eqnarray}
\delta\rho_a&=&M^{-1}_{ab}\delta\tilde{\mu}_b.
\end{eqnarray}
Thus, one can see that
\begin{eqnarray}
M&=&\chi^{-1},~~~\chi=M^{-1}.
\end{eqnarray}
Thus, to get the charge fluctuations, which were defined in the grand-canonical ensemble,  from a formulation based on the $f(T,\rho)$, one can proceed via,
\begin{eqnarray}
\tilde{\mu}_a&=&\frac{1}{T}\frac{\partial f}{\partial\rho_a},\\ \nonumber
\chi^{-1}_{ab}&=&\frac{1}{T}\frac{\partial^2 f}{\partial\rho_a\partial\rho_b}.
\end{eqnarray}
The charge fluctuation is then found by inverting $\chi^{-1}$. 

\subsubsection{The Specific Heat and Energy-Charge Fluctuation in the Canonical Ensemble}

One can also find the fluctuation of the energy density, at fixed $\tilde{\mu}$, in terms of $f(T,\rho)$.
\begin{eqnarray}
\delta\epsilon&=&
\frac{\partial\epsilon}{\partial\rho_a}\delta\rho_a
+\left.\frac{\partial\epsilon}{\partial T}\right|_{{\rm fixed}~\rho}\delta T,\\ \nonumber
\delta\tilde{\mu}_a&=&\frac{\partial\tilde{\mu_a}}{\partial\rho_b}\delta\rho_b+\left.\frac{\partial\tilde{\mu_a}}{\partial T}\right|_{{\rm fixed}~\rho}\delta T=0,\\ \nonumber
\delta\rho_a&=&-\left(\frac{\partial\tilde{\mu}}{\partial\rho}\right)_{ab}^{-1}\left.\frac{\partial\tilde{\mu}_b}{\partial T}\right|_{{\rm fixed}~\rho}\delta T,\\ \nonumber
&=&-\chi_{ab}\frac{\partial\tilde{\mu}_b}{\partial T}\delta T,\\ \nonumber
\delta\epsilon&=&\delta T\left\{
-\frac{\partial \epsilon}{\partial\rho_a}\chi_{ab}\left.\frac{\partial\tilde{\mu_b}}{\partial T}\right|_{{\rm fixed}~\rho}+\left.\frac{\partial\epsilon}{\partial T}\right|_{{\rm fixed}~\rho}
\right\}.
\end{eqnarray}
This yields
\begin{eqnarray}\label{eq:chiEE}
\chi_{EE}&=&\left.\frac{\partial\epsilon}{\partial T}\right|_{{\rm fixed}~\rho}
-\frac{\partial\epsilon}{\partial\rho_a}\chi_{ab}\left.\frac{\partial\tilde{\mu}_b}{\partial T}\right|_{{\rm fixed}~\rho}.
\end{eqnarray}
Here, even if one is calculationg a contribution from a subset of the free energy, e.g. that due to interactions, $\chi_{ab}$ is the total charge susceptibility. 

Next, one can calculate $\chi_{EQ}$. We do similar steps above, except set $\delta T=0$.
\begin{eqnarray}
\delta\epsilon&=&\frac{\partial\epsilon}{\partial\rho_a}\delta\rho_a,\\ \nonumber
\delta\rho_a&=&\chi_{ab}\delta\tilde{\mu}_b,\\ \nonumber
\delta\mu_a&=&\chi^{-1}_{ab}\delta\rho_b,\\ \nonumber
\label{eq:chiEQ}
\chi_{EQ}&=&\frac{\partial\epsilon}{\partial\tilde{\mu}_a}=\frac{\partial\epsilon}{\partial\rho_b}\chi_{ba}.
\end{eqnarray}
Again, even if one is calculating the contribution to some subset of the free energy, $\chi_{ab}$ in Eq. (\ref{eq:chiEQ})must refer to the total charge susceptibility.

\subsection{Adding Contributions from the Non-Interacting Gas and from Interactions}

Because the interaction energy is mainly a function of the density, it is natural to express the Helmholtz free-energy as two separate contributions, which then add together. The first contribution to the free energy, that from the non-interacting gas, can be calculated by beginning with the grand canonical ensemble. The contribution from interactions, can then be added. For non-momentum dependent interactions one expects the free-energy density due to interactions to depend only on the density, not the temperature, though the derivation here does not assume that. We then express how the charge susceptibilites can then be generated from the Helmholtz free-energy density. It should be emphasized that stating that the interaction contribution to $f(\rho,T)$ adds as separate term, is crucial to this derivation. One could have stated that the interactions provided a separate term to the pressure, with both terms evaluated at the same chemical potential. But, that represents a different physical assumption that seems more difficult to motivate given that it is more natural to assume that the interaction contributes as a function of the densities, not as a function of the chemical potentials. For example, if one added an additional non-interacting species, the chemical potentials would change, but the densities of the interacting particles would not. For mean-field interactions, one might expect that the energy, $E$, changes with density, mostly independent of temperature, and that the entropy is unaffected. That implies that $F=E-TS$ would be altered by adding some function of the density. Further, one may wish to understand the susceptibilities in the mixed-phase region, i.e. densities at which the grand-canonical ensemble gives undefined fluctuations.

For our aims, we need to have thermodynamic quantities as a function of density, not chemical potential. Going from one to the other is not particularly difficult, unless one has an unstable mixed-phase region. If one wishes to model the dynamics of such a phase it is imperative to work with quantities which are functions of density. Further, the interaction energy is more straight-forward to describe as an addition to the free-energy density, because it is more often more easily understood as a function of density. For that reason one might think it reasonable to express the free-energy density as
\begin{eqnarray}
f(\rho,T)&=&f_{\rm free}(\vec{\rho},T)+f_{\rm int}(\vec{\rho},T).
\end{eqnarray}
In fact, if the interaction is purely density dependent, i.e. independent of the momenta of the underlying particles, then $f_{\rm int}(\rho,T)$ is simply the potential energy density, and is independent of temperature.

Given the definition in Eq. (\ref{eq:f12}), one can immediately see that
\begin{eqnarray}
\tilde{\mu}_{\rm tot}(\vec{\rho},T)&=&\tilde{\mu}_{\rm free}(\vec{\rho},T)+\tilde{\mu}_{\rm int}(\vec{\rho},T),\\ \nonumber
P_{\rm tot}(\vec{\rho},T)&=&P_{\rm free}(\vec{\rho},T)+P_{\rm int}(\vec{\rho},T),\\ \nonumber
\epsilon_{\rm tot}(\vec{\rho},T)&=&\epsilon_{\rm free}(\vec{\rho},T)+\epsilon_{\rm int}(\vec{\rho},T).
\end{eqnarray}
Note that this definition of $P_{\rm int}$ differs from one where the total pressure is defined by separating out the two contributions at the same chemical potential. The latter choice would be natural if one were working in the grand canonical ensemble, in terms of $\vec{\tilde{\mu}}$, rather than in terms of the density.

Thus, give the choice of Eq. (\ref{eq:f12}) one can see that
\begin{eqnarray}
\frac{\partial^2f_{\rm tot}}{\partial\rho_a\partial\rho_b}&=&
\frac{\partial^2f_{\rm free}}{\partial\rho_a\partial\rho_b}+\frac{\partial^2f_{\rm int}}{\partial\rho_a\partial\rho_b},\\ \nonumber
\chi^{-1}_{\rm tot}&=&\chi^{-1}_{\rm free}+\chi^{-1}_{\rm int}.
\end{eqnarray}
This if one wants the charge susceptibility of the entire system, one should calculate $\chi^{-1}$ for both pieces, add them together, then take the inverse of the matrix.
\begin{eqnarray}
\chi&=&(\chi^{-1}_{\rm free}+\chi^{-1}_{\rm int})^{-1}.
\end{eqnarray}
The susceptibilities which include energy can also be found in terms of density-dependent (rather that $\tilde{\mu}-$dependent functions. Beginning by expressing the grand partition function in terms of $T$ and $\tilde{\mu}$, one can write the energy charge fluctuation,
\begin{eqnarray}
\zeta_a&\equiv&\frac{1}{VT}\langle\delta E\delta Q_a\rangle=\frac{1}{T}\frac{\partial\epsilon}{\partial\tilde{\tilde{\mu}}_a}
=T\frac{\partial\rho_a}{\partial T}.
\end{eqnarray}
One can then write this as
\begin{eqnarray}
\zeta_a&=&\frac{1}{T}\frac{\partial\epsilon}{\partial\rho_b}\frac{\partial\rho_b}{\partial\tilde{\mu}_a}\\ \nonumber
&=&\frac{1}{T}\chi_{ab}\frac{\partial\epsilon}{\partial\rho_b}.
\end{eqnarray}

Rewriting the specific heat requires remembering that derivatives w.r.t. temperature require noting whether $\vec{\tilde{\mu}}$ or $\vec{\rho}$ are being held constant. First, we define the specific heat so that it corresponds to the energy fluctuation in the grand canonical ensemble,
\begin{eqnarray}
\xi&\equiv&\frac{1}{T^2V}\langle\delta E\delta E\rangle\\ \nonumber
&=&\left.\frac{\partial\epsilon}{\partial T}\right|_{\vec{\tilde{\mu}}}.
\end{eqnarray}
Next, one can determine how $\vec{\tilde{\mu}}$ must be changing in order to maintain constant density,
\begin{eqnarray}
\delta\rho_a&=&0=\left.\frac{\partial\rho_a}{\partial T}\right|_{\vec{\tilde{\mu}}}+\frac{\partial\rho_a}{\partial\tilde{\mu}_b}\delta\tilde{\mu}_b,\\ \nonumber
\delta\tilde{\mu}_a&=&-\chi^{-1}_{ab}\left.\frac{\partial\rho_a}{\partial T}\right|_{\vec{\tilde{\mu}}}.
\end{eqnarray}
This then gives
\begin{eqnarray}
\delta\epsilon&=&\left.\frac{\partial\epsilon}{\partial T}\right|_{\vec{\tilde{\mu}}}+\frac{\partial\epsilon}{\partial\tilde{\mu}_a}\delta\tilde{\mu}_a\\ \nonumber
\left.\frac{\partial\epsilon}{\partial T}\right|_{\vec{\rho}}
&=&\left.\frac{\partial\epsilon}{\partial T}\right|_{\vec{\tilde{\mu}}}-\left.\frac{\partial\rho_a}{\partial T}\right|_{\vec\tilde{\mu}}\chi^{-1}_{ab}\frac{\partial\epsilon}{\partial\tilde{\mu}_b}\\ \nonumber
&=&\left.\frac{\partial\epsilon}{\partial T}\right|_{\vec{\tilde{\mu}}}-\zeta_a\chi^{-1}_{ab}\zeta_b.
\end{eqnarray}

For our calculation, we have two components, that of a non-interacting gas, plus an interaction. The thermodynamics of the non-interactiong gas are most easily expressed in terms of chemical potentials, whereas the interaction term is most easily expressed in terms of densities. To add quantities together, we assume that the net free energy is the sum of the free energy of a gas and the interaction part,
\begin{eqnarray}
f&=&f^{\rm(gas)}+f^{\rm(int)}.
\end{eqnarray}
This then implies that
\begin{eqnarray}
\epsilon&=&\epsilon^{\rm(gas)}+\epsilon^{\rm(int)},\\ \nonumber
\tilde{\mu}_a&=&\tilde{\mu}_a^{\rm(gas)}+\tilde{\mu}_b^{\rm(int)},\\ \nonumber
\chi^{-1}&=&\chi^{-1{\rm(gas)}}+\chi^{-1{\rm(int)}},\\ \nonumber
\left.\frac{\partial\epsilon}{\partial T}\right|_{\vec{\rho}}&=&
\left.\frac{\partial\epsilon^{\rm(gas)}}{\partial T}\right|_{\vec{\rho}}
+\left.\frac{\partial\epsilon^{\rm(int)}}{\partial T}\right|_{\vec{\rho}}.
\end{eqnarray}
Note that the chemical potentials of the two parts add (they are not the same) and that the densities are the same. 

The order of operations for adding the two contributions are:
\begin{enumerate}\itemsep=0pt
        \item Calculate all quantities for the gas in the grand canonical ensemble. This includes $\epsilon^{\rm(gas)}$, $\zeta_a^{\rm(gas)}$, and $\chi^{\rm(gas)}$. Also take the inverse of $\chi^{\rm(gas)}$ to get $\chi^{-1{\rm(gas)}}$. Also calculate
        \begin{eqnarray}
        \frac{\partial\epsilon^{\rm(gas)}}{\partial\rho_a}&=&\frac{1}{T}\chi^{-1{\rm(gas)}}_{ab}\zeta^{\rm(gas)}_b,\\ \nonumber
        \left.\frac{\partial\epsilon^{\rm(gas)}}{\partial T}\right|_{\vec{\rho}}&=&\left.\frac{\partial\epsilon^{\rm(gas)}}{\partial T}\right|_{\vec{\tilde{\mu}}}-\zeta^{\rm(gas)}_a\chi^{-1{\rm(gas)}}_{ab}\zeta^{\rm(gas)}_b,\\ \nonumber
        \end{eqnarray}
        \item Calculate the quantities for the interaction contribution: 
        \begin{eqnarray}
        \epsilon^{\rm(int)}&=&-T^2\frac{\partial}{\partial T}(f^{\rm(int)}/T),\\ \nonumber
        \tilde{\mu}_a^{\rm(int)}&=&\frac{1}{T}\frac{\partial f^{\rm(int)}}{\partial\rho_a},\\ \nonumber
        \chi^{-1{\rm(int)}}_{ab}&=&\frac{\partial\tilde{\mu}^{\rm(int)}_a}{\partial\rho_b}.
        \end{eqnarray}
        along with $\partial\epsilon^{\rm(gas)}/\partial\rho_a$ and $\partial\epsilon^{\rm(gas)}/\partial T|_{\vec{\rho}}$.
        \item Add contribution from gas and interactions to  get combined quantities,
        \begin{eqnarray}
        \epsilon&=&\epsilon^{\rm(gas)}+\epsilon^{\rm(int)},\\ \nonumber
        P&=&P^{\rm(gas)}+P^{\rm(int)},\\ \nonumber
        \tilde{\mu}_a&=&\tilde{\mu}_a^{\rm(gas)}+\tilde{\mu}_a^{\rm(int)},\\ \nonumber
        \chi^{-1}_{ab}&=&\chi^{-1{\rm(gas)}}_{ab}+\chi^{-1{\rm(int)}}_{ab},\\ \nonumber
        \frac{\partial\epsilon}{\partial\rho_a}&=&\frac{\partial\epsilon^{\rm(gas)}}{\partial\rho_a}+\frac{\partial\epsilon^{\rm(int)}}{\partial\rho_a},\\ \nonumber
        \left.\frac{\partial\epsilon}{\partial T}\right|_{\vec{\rho}}&=&\left.\frac{\partial\epsilon^{\rm(gas)}}{\partial T}\right|_{\vec{\rho}}
        +\left.\frac{\partial\epsilon^{\rm(int)}}{\partial T}\right|_{\vec{\rho}}.
        \end{eqnarray}
        Now, from all the combined quantities, one can calculate all the net susceptibilities,
        \begin{eqnarray}
        \chi_{ab}&=&\left(\chi^{-1}\right)^{-1}_{ab},\\ \nonumber
        \zeta_a&=&\frac{1}{T}\chi_{ab}\frac{\partial\epsilon}{\partial\rho_b},\\ \nonumber
        \xi&=&\left.\frac{\partial\epsilon}{\partial T}\right|_{\vec{\rho}}+\zeta_a\chi^{-1}_{ab}\zeta_b.
        \end{eqnarray}
\end{enumerate}
\subsection{The speed of sound}
When taking partial derivatives, assume functions are of $T$ and $\tilde{\mu}_a\equiv\mu_a/T$.
\begin{eqnarray}
c_s^2&=&\left.\frac{\partial P}{\partial\epsilon}\right|_{s/\rho},\\ \nonumber
\delta P&=&\frac{\partial P}{\partial T}\delta T+\frac{\partial P}{\partial\tilde{\mu}_a}\delta\tilde{\mu}_a,\\ \nonumber
&=&\frac{P+\epsilon}{T}\delta T+\rho_aT\delta\tilde{\mu}_a,\\ \nonumber
\delta \epsilon&=&\frac{\partial\epsilon}{\partial T}\delta T+\frac{\partial\epsilon}{\partial\tilde{\mu}_a}\delta\tilde{\mu}_a,\\ \nonumber
\delta\rho_a&=&\frac{\partial \rho_a}{\partial T}\delta T+\frac{\partial \rho_a}{\partial\tilde{\mu}_b}\delta\tilde{\mu}_b\\ \nonumber
&=&\frac{\partial \rho_a}{\partial T}\delta T+\chi_{ab}\delta\tilde{\mu}_b,\\ \nonumber
\delta s&=&\frac{\partial s}{\partial T}\delta T+\frac{\partial s}{\partial\tilde{\mu}_a}\delta\tilde{\mu}_a,\\ \nonumber
s^2\delta\left(\frac{\rho_a}{s}\right)&=&s\delta\rho_a-\rho_a\delta s=0,\\ \nonumber
0&=&s\frac{\partial\rho_a}{\partial T}\delta T+s\chi_{ab}\delta\tilde{\mu}_b
-\rho_b\frac{\partial s}{\partial T}\delta T-\rho_a\frac{\partial s}{\partial\tilde{\mu}_b}\delta\tilde{\mu}_b,\\ \nonumber
\delta\tilde{\mu}_a&=&M_a\frac{\delta T}{T},\\ \nonumber
\label{eq:Mdef}
M_a&=&T\left(s\chi-\rho\frac{\partial s}{\partial\tilde{\mu}}\right)_{ab}^{-1}
\left(\rho_b\frac{\partial s}{\partial T}-s\frac{\partial\rho_b}{\partial T}\right).
\end{eqnarray}
By substituting for $\delta\tilde{\mu}$ in the expressions for $\delta P$ and $\delta\epsilon$, then taking the ratio $\delta P/\delta\epsilon$, the factors of $\delta T$ drop out and
\begin{eqnarray}
c_s^2&=&\frac{(P+\epsilon)/T+\rho_aM_a}{\frac{\partial\epsilon}{\partial T}+\frac{\partial\epsilon}{\partial\tilde{\mu}_a}M_a/T}
\end{eqnarray}
Next, substitute to replace derivative of $s$, with derivatives of $\epsilon$ or $\rho$, so that all derivatives in the Eq. (\ref{eq:Mdef}) are related to fluctuations.
\begin{eqnarray}
\frac{\partial P}{\partial T}&=&\frac{P+\epsilon}{T},\\ \nonumber
\frac{\partial s}{\partial T}&=&\frac{\partial}{\partial T}\left(\frac{P+\epsilon-\tilde\mu_a\rho_aT}{T}\right)\\ \nonumber
&=&\frac{1}{T}\left(\frac{P+\epsilon}{T}+\frac{\partial\epsilon}{\partial T}-T\tilde{\mu_a}\frac{\partial\rho_a}{\partial T}-\tilde{\mu}_a\rho_a\right)-\frac{s}{T}\\ \nonumber
&=&\frac{1}{T}\frac{\partial\epsilon}{\partial T}-\tilde{\mu}_a\frac{\partial\rho_a}{\partial T},\\ \nonumber
\frac{\partial s}{\partial\tilde{\mu}_a}&=&\frac{1}{T}\frac{\partial}{\partial\tilde{\mu}_a}(P+\epsilon-\tilde{\mu}_b\rho_bT),\\ \nonumber
&=&\frac{1}{T}\left(
\rho_aT+\frac{\partial\epsilon}{\partial\tilde{\mu}_a}-\rho_aT-\tilde{\mu_b}T\frac{\partial\tilde{\rho}_b}{\partial\tilde{\mu}_a}
\right)\\ \nonumber
&=&\frac{1}{T}\left(
\frac{\partial\epsilon}{\partial\tilde{\mu}_a}-\tilde{\mu}_bT\frac{\partial\rho_b}{\partial\tilde{\mu}_a}
\right)\\ \nonumber
M_a&=&A^{-1}_{ab}\left(
\rho_b \frac{\partial\epsilon}{\partial T}  -\rho_bT\tilde{\mu}_c\frac{\partial\rho_c}{\partial T}-Ts\frac{\partial\rho_b}{\partial T}\right),\\ \nonumber
A_{ab}&=&s\chi_{ab}-\rho_a\left(
\frac{1}{T}\frac{\partial\epsilon}{\partial\tilde{\mu}_b}-\tilde{\mu}_c\frac{\partial\rho_c}{\partial\tilde{\mu}_b}
\right)\\ \nonumber
&=&s\chi_{ab}-\frac{\rho_a}{T}\frac{\partial\epsilon}{\partial\tilde{\mu}_b}+\rho_a\chi_{bc}\tilde{\mu}_c.
\end{eqnarray}
Now, let's summarize the results for the speed of sound using the following definitions for the susceptibilities,
\begin{eqnarray}
\xi&\equiv \frac{\partial\epsilon}{\partial T}=\frac{1}{T^2V}\langle\delta E\delta E\rangle,\\ \nonumber
\zeta_a&\equiv\frac{1}{T}\frac{\partial\epsilon}{\partial\tilde{\mu}_a}=T\frac{\partial\rho_a}{\partial T}
=\frac{1}{TV}\langle\delta E\delta Q_a\rangle,\\ \nonumber
\chi_{ab}&\equiv\frac{\partial\rho_a}{\partial\tilde{\mu}_b}=\frac{1}{V}\langle\delta Q_a\delta Q_b\rangle,\\ \nonumber
F_p&=&\frac{P+\epsilon}{T}=\frac{1}{T^2v}\langle\delta P_x\delta P_x\rangle,
\end{eqnarray}
where $P_x$ is a $x-$component of the momentum.

This gives
\begin{eqnarray}\label{eq:f12}
c_s^2&=&\frac{(P+\epsilon)/T+\rho_a M_a}{\xi+\zeta_aM_a},\\ \nonumber
M_a&=&A^{-1}_{ab}[(\xi-\tilde{\mu}_c\zeta_c)\rho_b-s\zeta_b],\\ \nonumber
A_{ab}&=&s\chi_{ab}-\rho_a\zeta_b.
\end{eqnarray}

\subsection{Derivatives at fixed energy density}

Stepping away from the topic of fluctuations, hydrodynamic codes often require taking derivatives of various thermodynamic quantities with respect to density at fixed temperature. Here, we summarize the relations that were required here. First, we review derivative of the pressure.
\begin{eqnarray}\label{eq:dPdrhoT}
\left.\frac{\partial P}{\partial\rho_a}\right|_T&=&
\frac{\partial P}{\partial\tilde{\mu}_b}\frac{\partial\tilde{\mu}_b}{\partial\rho_a}\\ \nonumber
&=&\chi^{-1}_{ab}\rho_bT,\\ \nonumber
\left.\frac{\partial\epsilon}{\partial\rho_a}\right|_T&=&\chi^{-1}_{ab}\frac{\partial\epsilon}{\partial\tilde{\mu}_b}\\ \nonumber
&=&\chi^{-1}_{ab}\zeta_bT=0,\\ \nonumber
\left.\delta P\right|_\rho&=&\left[\left.\frac{\partial P}{\partial T}\right|_{\tilde{\mu}}
-\left.\frac{\partial P}{\partial\tilde{\mu}_a}\right|_{\tilde{\mu}}\frac{1}{T}\chi^{-1}_{ab}\zeta_b
\right]\delta T,\\ \nonumber
\left.\frac{\partial P}{\partial T}\right|_\rho&=&\frac{P+\epsilon}{T}-\rho_a\chi^{-1}_{ab}\zeta_b.
\end{eqnarray}
Similarly, derivatives of $\epsilon$ are
\begin{eqnarray}\label{eq:dedrhoT}
\left.\frac{\partial\epsilon}{\partial T}\right|_\rho&=&
\left.\frac{\partial\epsilon}{\partial T}\right|_{\tilde{\mu}}
-\frac{\partial\epsilon}{\partial\tilde{\mu}_a}\frac{1}{T}\chi^{-1}_{ab}\zeta_b\\ \nonumber
&=&\xi-\zeta_a\chi^{-1}_{ab}\zeta_b.
\end{eqnarray}
Next, we need to find derivatives of $P$ and $\epsilon$ w.r.t $T$ at fixed density,
\begin{eqnarray}
\delta P&=&\frac{\partial P}{\partial T}\delta T+\frac{\partial P}{\partial\tilde{\mu}_a}\delta\tilde{\mu}_a,\\ \nonumber
\delta\rho_a&=&\frac{\partial\rho}{\partial T}\delta T+\frac{\partial\rho_a}{\partial\tilde{\mu}_b}\delta\tilde{\mu}_b=0\\ \nonumber
&=&\frac{1}{T}\zeta_a\delta T+\chi_{ab}\delta\tilde{\mu}_b
\end{eqnarray}
Now, to find $\partial P/\partial \rho_a|_\epsilon$,
\begin{eqnarray}
\delta P&=&\left.\frac{\partial P}{\partial\rho_a}\right|_T\delta\rho_a+\left.\frac{\partial P}{\partial T}\right|_\rho\delta T,\\ \nonumber
\delta \epsilon&=&\left.\frac{\partial \epsilon}{\partial\rho_a}\right|_T\delta\rho_a+\left.\frac{\partial \epsilon}{\partial T}\right|_\rho\delta T=0\\ \nonumber
\delta P&=&\left.\frac{\partial P}{\partial\rho_a}\right|_T\delta\rho_a-\frac{\delta\rho_a}{\partial\epsilon/\partial T|_\rho}
\left.\frac{\partial P}{\partial T}\right|_\rho\left.\frac{\partial\epsilon}{\partial\rho_a}\right|_T,\\ \nonumber
\left.\frac{\partial P}{\partial\rho_a}\right|_\epsilon&=&\left.\frac{\partial P}{\partial \rho_a}\right|_T
-\frac{1}{\xi}\left.\frac{\partial P}{\partial T}\right|_\rho\left.\frac{\partial\epsilon}{\partial\rho_a}\right|_T.
\end{eqnarray}
Finally, to find $\partial P/\partial\epsilon$ at fixed density, one can apply Eq.s (\ref{eq:dPdrhoT}) and (\ref{eq:dedrhoT}),
\begin{eqnarray}
\left.\frac{\partial P}{\partial\epsilon}\right|_\rho&=&\frac{\left.\frac{\partial P}{\partial T}\right|_\rho}
{\left.\frac{\partial\epsilon}{\partial T}\right|_\rho}
\end{eqnarray}

\subsection{Algorithmic Design}

The code used in this study has simple routines that calculate the free-particle contributions to $\vec{\rho},P,\epsilon$ and $\chi$ in terms of $\vec{\mu}$ and $T$. An optimized Newton's method routine is used to find $\vec{\mu}$ in terms of $\vec{\rho}$ and $T$. This results in an easily accessible routine that gives all the thermodyanmic quantities in terms of $\vec{\rho}$ and $T$. One can invert $\chi$ to find the free-energy contribution to $\chi^{-1}$. This thus provides $P_{\rm free}$, $\epsilon_{\rm free}, \vec{\mu}_{\rm free}$ and $\chi^{-1}_{\rm free}$.

Next, one calls a function that provides the free energy, plus all its derivatives with respect to $\vec{\rho}$. This provides the interaction contributions, $f_{\rm int}, P_{\rm int}$, $\epsilon_{\rm int}, \vec{\mu}_{\rm int}$ and $\chi^{-1}_{\rm int}$. If one has an analytic form for $f_{\rm int}(\vec{\rho},T)$ it is usually simple to find all the interaction contributions by taking derivatives of $f_{\rm int}$ with respect to temperature and densities. Finally, one finds the thermodynamic quantities of the entire system by simply adding the two pieces according to the steps outlined earlier.

\bibliography{refs.bib}

\begin{thebibliography}{44}%
\makeatletter
\providecommand \@ifxundefined [1]{%
 \@ifx{#1\undefined}
}%
\providecommand \@ifnum [1]{%
 \ifnum #1\expandafter \@firstoftwo
 \else \expandafter \@secondoftwo
 \fi
}%
\providecommand \@ifx [1]{%
 \ifx #1\expandafter \@firstoftwo
 \else \expandafter \@secondoftwo
 \fi
}%
\providecommand \natexlab [1]{#1}%
\providecommand \enquote  [1]{``#1''}%
\providecommand \bibnamefont  [1]{#1}%
\providecommand \bibfnamefont [1]{#1}%
\providecommand \citenamefont [1]{#1}%
\providecommand \href@noop [0]{\@secondoftwo}%
\providecommand \href [0]{\begingroup \@sanitize@url \@href}%
\providecommand \@href[1]{\@@startlink{#1}\@@href}%
\providecommand \@@href[1]{\endgroup#1\@@endlink}%
\providecommand \@sanitize@url [0]{\catcode `\\12\catcode `\$12\catcode
  `\&12\catcode `\#12\catcode `\^12\catcode `\_12\catcode `\%12\relax}%
\providecommand \@@startlink[1]{}%
\providecommand \@@endlink[0]{}%
\providecommand \url  [0]{\begingroup\@sanitize@url \@url }%
\providecommand \@url [1]{\endgroup\@href {#1}{\urlprefix }}%
\providecommand \urlprefix  [0]{URL }%
\providecommand \Eprint [0]{\href }%
\providecommand \doibase [0]{https://doi.org/}%
\providecommand \selectlanguage [0]{\@gobble}%
\providecommand \bibinfo  [0]{\@secondoftwo}%
\providecommand \bibfield  [0]{\@secondoftwo}%
\providecommand \translation [1]{[#1]}%
\providecommand \BibitemOpen [0]{}%
\providecommand \bibitemStop [0]{}%
\providecommand \bibitemNoStop [0]{.\EOS\space}%
\providecommand \EOS [0]{\spacefactor3000\relax}%
\providecommand \BibitemShut  [1]{\csname bibitem#1\endcsname}%
\let\auto@bib@innerbib\@empty
\bibitem [{\citenamefont {Busza}\ \emph {et~al.}(2018)\citenamefont {Busza},
  \citenamefont {Rajagopal},\ and\ \citenamefont {van~der
  Schee}}]{Busza:2018rrf}%
  \BibitemOpen
  \bibfield  {author} {\bibinfo {author} {\bibfnamefont {W.}~\bibnamefont
  {Busza}}, \bibinfo {author} {\bibfnamefont {K.}~\bibnamefont {Rajagopal}},\
  and\ \bibinfo {author} {\bibfnamefont {W.}~\bibnamefont {van~der Schee}},\
  }\bibfield  {title} {\bibinfo {title} {{Heavy Ion Collisions: The Big
  Picture, and the Big Questions}},\ }\href
  {https://doi.org/10.1146/annurev-nucl-101917-020852} {\bibfield  {journal}
  {\bibinfo  {journal} {Ann. Rev. Nucl. Part. Sci.}\ }\textbf {\bibinfo
  {volume} {68}},\ \bibinfo {pages} {339} (\bibinfo {year} {2018})},\ \Eprint
  {https://arxiv.org/abs/1802.04801} {arXiv:1802.04801 [hep-ph]} \BibitemShut
  {NoStop}%
\bibitem [{\citenamefont {Bzdak}\ \emph {et~al.}(2020)\citenamefont {Bzdak},
  \citenamefont {Esumi}, \citenamefont {Koch}, \citenamefont {Liao},
  \citenamefont {Stephanov},\ and\ \citenamefont {Xu}}]{Bzdak:2019pkr}%
  \BibitemOpen
  \bibfield  {author} {\bibinfo {author} {\bibfnamefont {A.}~\bibnamefont
  {Bzdak}}, \bibinfo {author} {\bibfnamefont {S.}~\bibnamefont {Esumi}},
  \bibinfo {author} {\bibfnamefont {V.}~\bibnamefont {Koch}}, \bibinfo {author}
  {\bibfnamefont {J.}~\bibnamefont {Liao}}, \bibinfo {author} {\bibfnamefont
  {M.}~\bibnamefont {Stephanov}},\ and\ \bibinfo {author} {\bibfnamefont
  {N.}~\bibnamefont {Xu}},\ }\bibfield  {title} {\bibinfo {title} {{Mapping the
  Phases of Quantum Chromodynamics with Beam Energy Scan}},\ }\href
  {https://doi.org/10.1016/j.physrep.2020.01.005} {\bibfield  {journal}
  {\bibinfo  {journal} {Phys. Rept.}\ }\textbf {\bibinfo {volume} {853}},\
  \bibinfo {pages} {1} (\bibinfo {year} {2020})},\ \Eprint
  {https://arxiv.org/abs/1906.00936} {arXiv:1906.00936 [nucl-th]} \BibitemShut
  {NoStop}%
\bibitem [{\citenamefont {Dietrich}\ \emph {et~al.}(2020)\citenamefont
  {Dietrich}, \citenamefont {Coughlin}, \citenamefont {Pang}, \citenamefont
  {Bulla}, \citenamefont {Heinzel}, \citenamefont {Issa}, \citenamefont
  {Tews},\ and\ \citenamefont {Antier}}]{Dietrich_2020}%
  \BibitemOpen
  \bibfield  {author} {\bibinfo {author} {\bibfnamefont {T.}~\bibnamefont
  {Dietrich}}, \bibinfo {author} {\bibfnamefont {M.~W.}\ \bibnamefont
  {Coughlin}}, \bibinfo {author} {\bibfnamefont {P.~T.~H.}\ \bibnamefont
  {Pang}}, \bibinfo {author} {\bibfnamefont {M.}~\bibnamefont {Bulla}},
  \bibinfo {author} {\bibfnamefont {J.}~\bibnamefont {Heinzel}}, \bibinfo
  {author} {\bibfnamefont {L.}~\bibnamefont {Issa}}, \bibinfo {author}
  {\bibfnamefont {I.}~\bibnamefont {Tews}},\ and\ \bibinfo {author}
  {\bibfnamefont {S.}~\bibnamefont {Antier}},\ }\bibfield  {title} {\bibinfo
  {title} {Multimessenger constraints on the neutron-star equation of state and
  the hubble constant},\ }\href {https://doi.org/10.1126/science.abb4317}
  {\bibfield  {journal} {\bibinfo  {journal} {Science}\ }\textbf {\bibinfo
  {volume} {370}},\ \bibinfo {pages} {1450} (\bibinfo {year}
  {2020})}\BibitemShut {NoStop}%
\bibitem [{\citenamefont {Beloin}\ \emph {et~al.}(2019)\citenamefont {Beloin},
  \citenamefont {Han}, \citenamefont {Steiner},\ and\ \citenamefont
  {Odbadrakh}}]{PhysRevC.100.055801}%
  \BibitemOpen
  \bibfield  {author} {\bibinfo {author} {\bibfnamefont {S.}~\bibnamefont
  {Beloin}}, \bibinfo {author} {\bibfnamefont {S.}~\bibnamefont {Han}},
  \bibinfo {author} {\bibfnamefont {A.~W.}\ \bibnamefont {Steiner}},\ and\
  \bibinfo {author} {\bibfnamefont {K.}~\bibnamefont {Odbadrakh}},\ }\bibfield
  {title} {\bibinfo {title} {Simultaneous fitting of neutron star structure and
  cooling data},\ }\href {https://doi.org/10.1103/PhysRevC.100.055801}
  {\bibfield  {journal} {\bibinfo  {journal} {Phys. Rev. C}\ }\textbf {\bibinfo
  {volume} {100}},\ \bibinfo {pages} {055801} (\bibinfo {year}
  {2019})}\BibitemShut {NoStop}%
\bibitem [{\citenamefont {Most}\ \emph {et~al.}(2020)\citenamefont {Most},
  \citenamefont {Papenfort}, \citenamefont {Dexheimer}, \citenamefont
  {Hanauske},\ and\ \citenamefont {Stoecker}}]{Hanauske2020a}%
  \BibitemOpen
  \bibfield  {author} {\bibinfo {author} {\bibfnamefont {E.~R.}\ \bibnamefont
  {Most}}, \bibinfo {author} {\bibfnamefont {L.~J.}\ \bibnamefont {Papenfort}},
  \bibinfo {author} {\bibfnamefont {V.}~\bibnamefont {Dexheimer}}, \bibinfo
  {author} {\bibfnamefont {M.}~\bibnamefont {Hanauske}},\ and\ \bibinfo
  {author} {\bibfnamefont {H.}~\bibnamefont {Stoecker}},\ }\href
  {https://doi.org/10.1140/epja/s10050-020-00073-4} {\bibfield  {journal}
  {\bibinfo  {journal} {Eur.~Phys.~J.~A}\ }\textbf {\bibinfo {volume} {56}},\
  \bibinfo {pages} {59} (\bibinfo {year} {2020})}\BibitemShut {NoStop}%
\bibitem [{\citenamefont {Weih}\ \emph {et~al.}(2020)\citenamefont {Weih},
  \citenamefont {Hanauske},\ and\ \citenamefont {Rezzolla}}]{Hanauske2020b}%
  \BibitemOpen
  \bibfield  {author} {\bibinfo {author} {\bibfnamefont {L.~R.}\ \bibnamefont
  {Weih}}, \bibinfo {author} {\bibfnamefont {M.}~\bibnamefont {Hanauske}},\
  and\ \bibinfo {author} {\bibfnamefont {L.}~\bibnamefont {Rezzolla}},\ }\href
  {https://doi.org/10.1103/PhysRevLett.124.171103} {\bibfield  {journal}
  {\bibinfo  {journal} {Phys. Rev. Lett.}\ }\textbf {\bibinfo {volume} {124}},\
  \bibinfo {pages} {171103} (\bibinfo {year} {2020})}\BibitemShut {NoStop}%
\bibitem [{\citenamefont {Most}\ \emph {et~al.}(2018)\citenamefont {Most},
  \citenamefont {Weih}, \citenamefont {Rezzolla},\ and\ \citenamefont
  {Schaffner-Bielich}}]{Schaffner2018}%
  \BibitemOpen
  \bibfield  {author} {\bibinfo {author} {\bibfnamefont {E.~R.}\ \bibnamefont
  {Most}}, \bibinfo {author} {\bibfnamefont {L.~R.}\ \bibnamefont {Weih}},
  \bibinfo {author} {\bibfnamefont {L.}~\bibnamefont {Rezzolla}},\ and\
  \bibinfo {author} {\bibfnamefont {J.}~\bibnamefont {Schaffner-Bielich}},\
  }\href {https://doi.org/10.1103/PhysRevLett.120.261103} {\bibfield  {journal}
  {\bibinfo  {journal} {Phys. Rev. Lett.}\ }\textbf {\bibinfo {volume} {120}},\
  \bibinfo {pages} {261103} (\bibinfo {year} {2018})}\BibitemShut {NoStop}%
\bibitem [{\citenamefont {Legred}\ \emph {et~al.}(2021)\citenamefont {Legred},
  \citenamefont {Chatziioannou}, \citenamefont {Essick}, \citenamefont {Han},\
  and\ \citenamefont {Landry}}]{PhysRevD.104.063003}%
  \BibitemOpen
  \bibfield  {author} {\bibinfo {author} {\bibfnamefont {I.}~\bibnamefont
  {Legred}}, \bibinfo {author} {\bibfnamefont {K.}~\bibnamefont
  {Chatziioannou}}, \bibinfo {author} {\bibfnamefont {R.}~\bibnamefont
  {Essick}}, \bibinfo {author} {\bibfnamefont {S.}~\bibnamefont {Han}},\ and\
  \bibinfo {author} {\bibfnamefont {P.}~\bibnamefont {Landry}},\ }\bibfield
  {title} {\bibinfo {title} {Impact of the psr $\mathrm{J}0740+6620$ radius
  constraint on the properties of high-density matter},\ }\href
  {https://doi.org/10.1103/PhysRevD.104.063003} {\bibfield  {journal} {\bibinfo
   {journal} {Phys. Rev. D}\ }\textbf {\bibinfo {volume} {104}},\ \bibinfo
  {pages} {063003} (\bibinfo {year} {2021})}\BibitemShut {NoStop}%
\bibitem [{\citenamefont {Drischler}\ \emph {et~al.}(2021)\citenamefont
  {Drischler}, \citenamefont {Han}, \citenamefont {Lattimer}, \citenamefont
  {Prakash}, \citenamefont {Reddy},\ and\ \citenamefont
  {Zhao}}]{Drischler_2021}%
  \BibitemOpen
  \bibfield  {author} {\bibinfo {author} {\bibfnamefont {C.}~\bibnamefont
  {Drischler}}, \bibinfo {author} {\bibfnamefont {S.}~\bibnamefont {Han}},
  \bibinfo {author} {\bibfnamefont {J.~M.}\ \bibnamefont {Lattimer}}, \bibinfo
  {author} {\bibfnamefont {M.}~\bibnamefont {Prakash}}, \bibinfo {author}
  {\bibfnamefont {S.}~\bibnamefont {Reddy}},\ and\ \bibinfo {author}
  {\bibfnamefont {T.}~\bibnamefont {Zhao}},\ }\bibfield  {title} {\bibinfo
  {title} {Limiting masses and radii of neutron stars and their implications},\
  }\href {https://doi.org/10.1103/physrevc.103.045808} {\bibfield  {journal}
  {\bibinfo  {journal} {Phys. Rev. C}\ }\textbf {\bibinfo {volume} {103}},\
  \bibinfo {pages} {045805} (\bibinfo {year} {2021})}\BibitemShut {NoStop}%
\bibitem [{\citenamefont {Huth}\ \emph {et~al.}(2022)\citenamefont {Huth},
  \citenamefont {Pang}, \citenamefont {Tews}, \citenamefont {Dietrich},
  \citenamefont {F{\`{e}}vre}, \citenamefont {Schwenk}, \citenamefont
  {Trautmann}, \citenamefont {Agarwal}, \citenamefont {Bulla}, \citenamefont
  {Coughlin},\ and\ \citenamefont {Broeck}}]{Huth_2022}%
  \BibitemOpen
  \bibfield  {author} {\bibinfo {author} {\bibfnamefont {S.}~\bibnamefont
  {Huth}}, \bibinfo {author} {\bibfnamefont {P.~T.~H.}\ \bibnamefont {Pang}},
  \bibinfo {author} {\bibfnamefont {I.}~\bibnamefont {Tews}}, \bibinfo {author}
  {\bibfnamefont {T.}~\bibnamefont {Dietrich}}, \bibinfo {author}
  {\bibfnamefont {A.~L.}\ \bibnamefont {F{\`{e}}vre}}, \bibinfo {author}
  {\bibfnamefont {A.}~\bibnamefont {Schwenk}}, \bibinfo {author} {\bibfnamefont
  {W.}~\bibnamefont {Trautmann}}, \bibinfo {author} {\bibfnamefont
  {K.}~\bibnamefont {Agarwal}}, \bibinfo {author} {\bibfnamefont
  {M.}~\bibnamefont {Bulla}}, \bibinfo {author} {\bibfnamefont {M.~W.}\
  \bibnamefont {Coughlin}},\ and\ \bibinfo {author} {\bibfnamefont {C.~V.~D.}\
  \bibnamefont {Broeck}},\ }\bibfield  {title} {\bibinfo {title} {Constraining
  neutron-star matter with microscopic and macroscopic collisions},\ }\href
  {https://doi.org/10.1038/s41586-022-04750-w} {\bibfield  {journal} {\bibinfo
  {journal} {Nature}\ }\textbf {\bibinfo {volume} {606}},\ \bibinfo {pages}
  {276} (\bibinfo {year} {2022})}\BibitemShut {NoStop}%
\bibitem [{\citenamefont {Neill}\ \emph {et~al.}()\citenamefont {Neill},
  \citenamefont {Preston}, \citenamefont {Newton},\ and\ \citenamefont
  {Tsang}}]{https://doi.org/10.48550/arxiv.2208.00994}%
  \BibitemOpen
  \bibfield  {author} {\bibinfo {author} {\bibfnamefont {D.}~\bibnamefont
  {Neill}}, \bibinfo {author} {\bibfnamefont {R.}~\bibnamefont {Preston}},
  \bibinfo {author} {\bibfnamefont {W.~G.}\ \bibnamefont {Newton}},\ and\
  \bibinfo {author} {\bibfnamefont {D.}~\bibnamefont {Tsang}},\ }\href
  {https://doi.org/10.48550/ARXIV.2208.00994} {}\Eprint
  {https://arxiv.org/abs/2208.00994} {arXiv:2208.00994 [astro-ph.HE]}
  \BibitemShut {NoStop}%
\bibitem [{\citenamefont {Sorensen}\ \emph
  {et~al.}(2023{\natexlab{a}})\citenamefont {Sorensen} \emph
  {et~al.}}]{Sorensen:2023zkk}%
  \BibitemOpen
  \bibfield  {author} {\bibinfo {author} {\bibfnamefont {A.}~\bibnamefont
  {Sorensen}} \emph {et~al.},\ }\href@noop {} {\bibinfo {title} {Dense nuclear
  matter equation of state from heavy-ion collisions}} (\bibinfo {year}
  {2023}{\natexlab{a}}),\ \Eprint {https://arxiv.org/abs/2301.13253}
  {arXiv:2301.13253 [nucl-th]} \BibitemShut {NoStop}%
\bibitem [{\citenamefont {Hebeler}\ \emph {et~al.}(2010)\citenamefont
  {Hebeler}, \citenamefont {Lattimer}, \citenamefont {Pethick},\ and\
  \citenamefont {Schwenk}}]{PhysRevLett.105.161102}%
  \BibitemOpen
  \bibfield  {author} {\bibinfo {author} {\bibfnamefont {K.}~\bibnamefont
  {Hebeler}}, \bibinfo {author} {\bibfnamefont {J.~M.}\ \bibnamefont
  {Lattimer}}, \bibinfo {author} {\bibfnamefont {C.~J.}\ \bibnamefont
  {Pethick}},\ and\ \bibinfo {author} {\bibfnamefont {A.}~\bibnamefont
  {Schwenk}},\ }\bibfield  {title} {\bibinfo {title} {Constraints on neutron
  star radii based on chiral effective field theory interactions},\ }\href
  {https://doi.org/10.1103/PhysRevLett.105.161102} {\bibfield  {journal}
  {\bibinfo  {journal} {Phys. Rev. Lett.}\ }\textbf {\bibinfo {volume} {105}},\
  \bibinfo {pages} {161102} (\bibinfo {year} {2010})}\BibitemShut {NoStop}%
\bibitem [{\citenamefont {Gandolfi}\ \emph {et~al.}(2012)\citenamefont
  {Gandolfi}, \citenamefont {Carlson},\ and\ \citenamefont
  {Reddy}}]{PhysRevC.85.032801}%
  \BibitemOpen
  \bibfield  {author} {\bibinfo {author} {\bibfnamefont {S.}~\bibnamefont
  {Gandolfi}}, \bibinfo {author} {\bibfnamefont {J.}~\bibnamefont {Carlson}},\
  and\ \bibinfo {author} {\bibfnamefont {S.}~\bibnamefont {Reddy}},\ }\bibfield
   {title} {\bibinfo {title} {Maximum mass and radius of neutron stars, and the
  nuclear symmetry energy},\ }\href
  {https://doi.org/10.1103/PhysRevC.85.032801} {\bibfield  {journal} {\bibinfo
  {journal} {Phys. Rev. C}\ }\textbf {\bibinfo {volume} {85}},\ \bibinfo
  {pages} {032801} (\bibinfo {year} {2012})}\BibitemShut {NoStop}%
\bibitem [{\citenamefont {Drischler}\ \emph {et~al.}(2014)\citenamefont
  {Drischler}, \citenamefont {Som\`a},\ and\ \citenamefont
  {Schwenk}}]{PhysRevC.89.025806}%
  \BibitemOpen
  \bibfield  {author} {\bibinfo {author} {\bibfnamefont {C.}~\bibnamefont
  {Drischler}}, \bibinfo {author} {\bibfnamefont {V.}~\bibnamefont {Som\`a}},\
  and\ \bibinfo {author} {\bibfnamefont {A.}~\bibnamefont {Schwenk}},\
  }\bibfield  {title} {\bibinfo {title} {Microscopic calculations and energy
  expansions for neutron-rich matter},\ }\href
  {https://doi.org/10.1103/PhysRevC.89.025806} {\bibfield  {journal} {\bibinfo
  {journal} {Phys. Rev. C}\ }\textbf {\bibinfo {volume} {89}},\ \bibinfo
  {pages} {025806} (\bibinfo {year} {2014})}\BibitemShut {NoStop}%
\bibitem [{\citenamefont {Steinheimer}\ and\ \citenamefont
  {Randrup}(2012)}]{Steinheimer:2012gc}%
  \BibitemOpen
  \bibfield  {author} {\bibinfo {author} {\bibfnamefont {J.}~\bibnamefont
  {Steinheimer}}\ and\ \bibinfo {author} {\bibfnamefont {J.}~\bibnamefont
  {Randrup}},\ }\bibfield  {title} {\bibinfo {title} {{Spinodal amplification
  of density fluctuations in fluid-dynamical simulations of relativistic
  nuclear collisions}},\ }\href
  {https://doi.org/10.1103/PhysRevLett.109.212301} {\bibfield  {journal}
  {\bibinfo  {journal} {Phys. Rev. Lett.}\ }\textbf {\bibinfo {volume} {109}},\
  \bibinfo {pages} {212301} (\bibinfo {year} {2012})},\ \Eprint
  {https://arxiv.org/abs/1209.2462} {arXiv:1209.2462 [nucl-th]} \BibitemShut
  {NoStop}%
\bibitem [{\citenamefont {Steinheimer}\ and\ \citenamefont
  {Randrup}(2013)}]{Steinheimer:2013glaa}%
  \BibitemOpen
  \bibfield  {author} {\bibinfo {author} {\bibfnamefont {J.}~\bibnamefont
  {Steinheimer}}\ and\ \bibinfo {author} {\bibfnamefont {J.}~\bibnamefont
  {Randrup}},\ }\bibfield  {title} {\bibinfo {title} {{Spinodal density
  enhancements in simulations of relativistic nuclear collisions}},\ }\href
  {https://doi.org/10.1103/PhysRevC.87.054903} {\bibfield  {journal} {\bibinfo
  {journal} {Phys. Rev. C}\ }\textbf {\bibinfo {volume} {87}},\ \bibinfo
  {pages} {054903} (\bibinfo {year} {2013})},\ \Eprint
  {https://arxiv.org/abs/1302.2956} {arXiv:1302.2956 [nucl-th]} \BibitemShut
  {NoStop}%
\bibitem [{\citenamefont {Savchuk}\ \emph {et~al.}(2023)\citenamefont
  {Savchuk}, \citenamefont {Poberezhnyuk}, \citenamefont {Motornenko},
  \citenamefont {Steinheimer}, \citenamefont {Gorenstein},\ and\ \citenamefont
  {Vovchenko}}]{Savchuk:2022msa}%
  \BibitemOpen
  \bibfield  {author} {\bibinfo {author} {\bibfnamefont {O.}~\bibnamefont
  {Savchuk}}, \bibinfo {author} {\bibfnamefont {R.~V.}\ \bibnamefont
  {Poberezhnyuk}}, \bibinfo {author} {\bibfnamefont {A.}~\bibnamefont
  {Motornenko}}, \bibinfo {author} {\bibfnamefont {J.}~\bibnamefont
  {Steinheimer}}, \bibinfo {author} {\bibfnamefont {M.~I.}\ \bibnamefont
  {Gorenstein}},\ and\ \bibinfo {author} {\bibfnamefont {V.}~\bibnamefont
  {Vovchenko}},\ }\bibfield  {title} {\bibinfo {title} {{Phase transition
  amplification of proton number fluctuations in nuclear collisions from a
  transport model approach}},\ }\href
  {https://doi.org/10.1103/PhysRevC.107.024913} {\bibfield  {journal} {\bibinfo
   {journal} {Phys. Rev. C}\ }\textbf {\bibinfo {volume} {107}},\ \bibinfo
  {pages} {024913} (\bibinfo {year} {2023})},\ \Eprint
  {https://arxiv.org/abs/2211.13200} {arXiv:2211.13200 [hep-ph]} \BibitemShut
  {NoStop}%
\bibitem [{\citenamefont {Vovchenko}\ \emph {et~al.}(2020)\citenamefont
  {Vovchenko}, \citenamefont {Savchuk}, \citenamefont {Poberezhnyuk},
  \citenamefont {Gorenstein},\ and\ \citenamefont {Koch}}]{Vovchenko:2020tsr}%
  \BibitemOpen
  \bibfield  {author} {\bibinfo {author} {\bibfnamefont {V.}~\bibnamefont
  {Vovchenko}}, \bibinfo {author} {\bibfnamefont {O.}~\bibnamefont {Savchuk}},
  \bibinfo {author} {\bibfnamefont {R.~V.}\ \bibnamefont {Poberezhnyuk}},
  \bibinfo {author} {\bibfnamefont {M.~I.}\ \bibnamefont {Gorenstein}},\ and\
  \bibinfo {author} {\bibfnamefont {V.}~\bibnamefont {Koch}},\ }\bibfield
  {title} {\bibinfo {title} {{Connecting fluctuation measurements in heavy-ion
  collisions with the grand-canonical susceptibilities}},\ }\href
  {https://doi.org/10.1016/j.physletb.2020.135868} {\bibfield  {journal}
  {\bibinfo  {journal} {Phys. Lett. B}\ }\textbf {\bibinfo {volume} {811}},\
  \bibinfo {pages} {135868} (\bibinfo {year} {2020})},\ \Eprint
  {https://arxiv.org/abs/2003.13905} {arXiv:2003.13905 [hep-ph]} \BibitemShut
  {NoStop}%
\bibitem [{\citenamefont {Sorensen}\ \emph
  {et~al.}(2023{\natexlab{b}})\citenamefont {Sorensen}, \citenamefont
  {Oliinychenko}, \citenamefont {McLerran},\ and\ \citenamefont
  {Koch}}]{Sorensen:2022odd}%
  \BibitemOpen
  \bibfield  {author} {\bibinfo {author} {\bibfnamefont {A.}~\bibnamefont
  {Sorensen}}, \bibinfo {author} {\bibfnamefont {D.}~\bibnamefont
  {Oliinychenko}}, \bibinfo {author} {\bibfnamefont {L.}~\bibnamefont
  {McLerran}},\ and\ \bibinfo {author} {\bibfnamefont {V.}~\bibnamefont
  {Koch}},\ }\bibfield  {title} {\bibinfo {title} {{Measuring the Speed of
  Sound Using Cumulants of Baryon Number}},\ }\href
  {https://doi.org/10.5506/APhysPolBSupp.16.1-A48} {\bibfield  {journal}
  {\bibinfo  {journal} {Acta Phys. Polon. Supp.}\ }\textbf {\bibinfo {volume}
  {16}},\ \bibinfo {pages} {1} (\bibinfo {year} {2023}{\natexlab{b}})},\
  \Eprint {https://arxiv.org/abs/2209.04957} {arXiv:2209.04957 [nucl-th]}
  \BibitemShut {NoStop}%
\bibitem [{\citenamefont {Ling}\ \emph {et~al.}(2014)\citenamefont {Ling},
  \citenamefont {Springer},\ and\ \citenamefont {Stephanov}}]{Ling:2013ksb}%
  \BibitemOpen
  \bibfield  {author} {\bibinfo {author} {\bibfnamefont {B.}~\bibnamefont
  {Ling}}, \bibinfo {author} {\bibfnamefont {T.}~\bibnamefont {Springer}},\
  and\ \bibinfo {author} {\bibfnamefont {M.}~\bibnamefont {Stephanov}},\
  }\bibfield  {title} {\bibinfo {title} {{Hydrodynamics of charge fluctuations
  and balance functions}},\ }\href {https://doi.org/10.1103/PhysRevC.89.064901}
  {\bibfield  {journal} {\bibinfo  {journal} {Phys. Rev. C}\ }\textbf {\bibinfo
  {volume} {89}},\ \bibinfo {pages} {064901} (\bibinfo {year} {2014})},\
  \Eprint {https://arxiv.org/abs/1310.6036} {arXiv:1310.6036 [nucl-th]}
  \BibitemShut {NoStop}%
\bibitem [{\citenamefont {Bass}\ \emph {et~al.}(2000)\citenamefont {Bass},
  \citenamefont {Danielewicz},\ and\ \citenamefont {Pratt}}]{Bass:2000az}%
  \BibitemOpen
  \bibfield  {author} {\bibinfo {author} {\bibfnamefont {S.~A.}\ \bibnamefont
  {Bass}}, \bibinfo {author} {\bibfnamefont {P.}~\bibnamefont {Danielewicz}},\
  and\ \bibinfo {author} {\bibfnamefont {S.}~\bibnamefont {Pratt}},\ }\bibfield
   {title} {\bibinfo {title} {{Clocking hadronization in relativistic heavy ion
  collisions with balance functions}},\ }\href
  {https://doi.org/10.1103/PhysRevLett.85.2689} {\bibfield  {journal} {\bibinfo
   {journal} {Phys. Rev. Lett.}\ }\textbf {\bibinfo {volume} {85}},\ \bibinfo
  {pages} {2689} (\bibinfo {year} {2000})},\ \Eprint
  {https://arxiv.org/abs/nucl-th/0005044} {arXiv:nucl-th/0005044} \BibitemShut
  {NoStop}%
\bibitem [{\citenamefont {Pratt}\ \emph {et~al.}(2015)\citenamefont {Pratt},
  \citenamefont {McCormack},\ and\ \citenamefont {Ratti}}]{Pratt:2015jsa}%
  \BibitemOpen
  \bibfield  {author} {\bibinfo {author} {\bibfnamefont {S.}~\bibnamefont
  {Pratt}}, \bibinfo {author} {\bibfnamefont {W.~P.}\ \bibnamefont
  {McCormack}},\ and\ \bibinfo {author} {\bibfnamefont {C.}~\bibnamefont
  {Ratti}},\ }\bibfield  {title} {\bibinfo {title} {{Production of Charge in
  Heavy Ion Collisions}},\ }\href {https://doi.org/10.1103/PhysRevC.92.064905}
  {\bibfield  {journal} {\bibinfo  {journal} {Phys. Rev. C}\ }\textbf {\bibinfo
  {volume} {92}},\ \bibinfo {pages} {064905} (\bibinfo {year} {2015})},\
  \Eprint {https://arxiv.org/abs/1508.07031} {arXiv:1508.07031 [nucl-th]}
  \BibitemShut {NoStop}%
\bibitem [{\citenamefont {Pratt}\ and\ \citenamefont
  {Young}(2017)}]{Pratt:2016lol}%
  \BibitemOpen
  \bibfield  {author} {\bibinfo {author} {\bibfnamefont {S.}~\bibnamefont
  {Pratt}}\ and\ \bibinfo {author} {\bibfnamefont {C.}~\bibnamefont {Young}},\
  }\bibfield  {title} {\bibinfo {title} {{Relating Measurable Correlations in
  Heavy Ion Collisions to Bulk Properties of Equilibrated QCD Matter}},\ }\href
  {https://doi.org/10.1103/PhysRevC.95.054901} {\bibfield  {journal} {\bibinfo
  {journal} {Phys. Rev. C}\ }\textbf {\bibinfo {volume} {95}},\ \bibinfo
  {pages} {054901} (\bibinfo {year} {2017})},\ \Eprint
  {https://arxiv.org/abs/1612.08983} {arXiv:1612.08983 [nucl-th]} \BibitemShut
  {NoStop}%
\bibitem [{\citenamefont {Pratt}\ and\ \citenamefont
  {Plumberg}(2019)}]{Pratt:2018ebf}%
  \BibitemOpen
  \bibfield  {author} {\bibinfo {author} {\bibfnamefont {S.}~\bibnamefont
  {Pratt}}\ and\ \bibinfo {author} {\bibfnamefont {C.}~\bibnamefont
  {Plumberg}},\ }\bibfield  {title} {\bibinfo {title} {{Evolving Charge
  Correlations in a Hybrid Model with both Hydrodynamics and Hadronic Boltzmann
  Descriptions}},\ }\href {https://doi.org/10.1103/PhysRevC.99.044916}
  {\bibfield  {journal} {\bibinfo  {journal} {Phys. Rev. C}\ }\textbf {\bibinfo
  {volume} {99}},\ \bibinfo {pages} {044916} (\bibinfo {year} {2019})},\
  \Eprint {https://arxiv.org/abs/1812.05649} {arXiv:1812.05649 [nucl-th]}
  \BibitemShut {NoStop}%
\bibitem [{\citenamefont {Pratt}\ and\ \citenamefont
  {Plumberg}(2021)}]{Pratt:2021xvg}%
  \BibitemOpen
  \bibfield  {author} {\bibinfo {author} {\bibfnamefont {S.}~\bibnamefont
  {Pratt}}\ and\ \bibinfo {author} {\bibfnamefont {C.}~\bibnamefont
  {Plumberg}},\ }\bibfield  {title} {\bibinfo {title} {{Charge balance
  functions for heavy-ion collisions at energies available at the CERN Large
  Hadron Collider}},\ }\href {https://doi.org/10.1103/PhysRevC.104.014906}
  {\bibfield  {journal} {\bibinfo  {journal} {Phys. Rev. C}\ }\textbf {\bibinfo
  {volume} {104}},\ \bibinfo {pages} {014906} (\bibinfo {year} {2021})},\
  \Eprint {https://arxiv.org/abs/2104.00628} {arXiv:2104.00628 [nucl-th]}
  \BibitemShut {NoStop}%
\bibitem [{\citenamefont {Aarts}\ \emph {et~al.}(2015)\citenamefont {Aarts},
  \citenamefont {Allton}, \citenamefont {Amato}, \citenamefont {Giudice},
  \citenamefont {Hands},\ and\ \citenamefont {Skullerud}}]{Aarts:2014nba}%
  \BibitemOpen
  \bibfield  {author} {\bibinfo {author} {\bibfnamefont {G.}~\bibnamefont
  {Aarts}}, \bibinfo {author} {\bibfnamefont {C.}~\bibnamefont {Allton}},
  \bibinfo {author} {\bibfnamefont {A.}~\bibnamefont {Amato}}, \bibinfo
  {author} {\bibfnamefont {P.}~\bibnamefont {Giudice}}, \bibinfo {author}
  {\bibfnamefont {S.}~\bibnamefont {Hands}},\ and\ \bibinfo {author}
  {\bibfnamefont {J.-I.}\ \bibnamefont {Skullerud}},\ }\bibfield  {title}
  {\bibinfo {title} {{Electrical conductivity and charge diffusion in thermal
  QCD from the lattice}},\ }\href {https://doi.org/10.1007/JHEP02(2015)186}
  {\bibfield  {journal} {\bibinfo  {journal} {JHEP}\ }\textbf {\bibinfo
  {volume} {02}},\ \bibinfo {pages} {186}},\ \Eprint
  {https://arxiv.org/abs/1412.6411} {arXiv:1412.6411 [hep-lat]} \BibitemShut
  {NoStop}%
\bibitem [{\citenamefont {Bjorken}(1983)}]{Bjorken:1982qr}%
  \BibitemOpen
  \bibfield  {author} {\bibinfo {author} {\bibfnamefont {J.~D.}\ \bibnamefont
  {Bjorken}},\ }\bibfield  {title} {\bibinfo {title} {{Highly Relativistic
  Nucleus-Nucleus Collisions: The Central Rapidity Region}},\ }\href
  {https://doi.org/10.1103/PhysRevD.27.140} {\bibfield  {journal} {\bibinfo
  {journal} {Phys. Rev. D}\ }\textbf {\bibinfo {volume} {27}},\ \bibinfo
  {pages} {140} (\bibinfo {year} {1983})}\BibitemShut {NoStop}%
\bibitem [{\citenamefont {Romatschke}(2010)}]{Romatschke:2010}%
  \BibitemOpen
  \bibfield  {author} {\bibinfo {author} {\bibfnamefont {P.}~\bibnamefont
  {Romatschke}},\ }\bibfield  {title} {\bibinfo {title} {{New Developments in
  Relativistic Viscous Hydrodynamics}},\ }\href
  {https://doi.org/10.1142/S0218301310014613} {\bibfield  {journal} {\bibinfo
  {journal} {Int. J. Mod. Phys. E}\ }\textbf {\bibinfo {volume} {19}},\
  \bibinfo {pages} {1} (\bibinfo {year} {2010})},\ \Eprint
  {https://arxiv.org/abs/0902.3663} {arXiv:0902.3663 [hep-ph]} \BibitemShut
  {NoStop}%
\bibitem [{\citenamefont {Pratt}\ \emph {et~al.}(2018)\citenamefont {Pratt},
  \citenamefont {Kim},\ and\ \citenamefont {Plumberg}}]{Pratt:2017oyf}%
  \BibitemOpen
  \bibfield  {author} {\bibinfo {author} {\bibfnamefont {S.}~\bibnamefont
  {Pratt}}, \bibinfo {author} {\bibfnamefont {J.}~\bibnamefont {Kim}},\ and\
  \bibinfo {author} {\bibfnamefont {C.}~\bibnamefont {Plumberg}},\ }\bibfield
  {title} {\bibinfo {title} {{Evolution of Charge Fluctuations and Correlations
  in the Hydrodynamic Stage of Heavy Ion Collisions}},\ }\href
  {https://doi.org/10.1103/PhysRevC.98.014904} {\bibfield  {journal} {\bibinfo
  {journal} {Phys. Rev. C}\ }\textbf {\bibinfo {volume} {98}},\ \bibinfo
  {pages} {014904} (\bibinfo {year} {2018})},\ \Eprint
  {https://arxiv.org/abs/1712.09298} {arXiv:1712.09298 [nucl-th]} \BibitemShut
  {NoStop}%
\bibitem [{\citenamefont {Pradeep}\ \emph {et~al.}(2022)\citenamefont
  {Pradeep}, \citenamefont {Rajagopal}, \citenamefont {Stephanov},\ and\
  \citenamefont {Yin}}]{Pradeep:2022mkf}%
  \BibitemOpen
  \bibfield  {author} {\bibinfo {author} {\bibfnamefont {M.}~\bibnamefont
  {Pradeep}}, \bibinfo {author} {\bibfnamefont {K.}~\bibnamefont {Rajagopal}},
  \bibinfo {author} {\bibfnamefont {M.}~\bibnamefont {Stephanov}},\ and\
  \bibinfo {author} {\bibfnamefont {Y.}~\bibnamefont {Yin}},\ }\bibfield
  {title} {\bibinfo {title} {{Freezing out fluctuations in Hydro+ near the QCD
  critical point}},\ }\href {https://doi.org/10.1103/PhysRevD.106.036017}
  {\bibfield  {journal} {\bibinfo  {journal} {Phys. Rev. D}\ }\textbf {\bibinfo
  {volume} {106}},\ \bibinfo {pages} {036017} (\bibinfo {year} {2022})},\
  \Eprint {https://arxiv.org/abs/2204.00639} {arXiv:2204.00639 [hep-ph]}
  \BibitemShut {NoStop}%
\bibitem [{\citenamefont {Pratt}\ and\ \citenamefont
  {Torrieri}(2010)}]{Pratt:2010jt}%
  \BibitemOpen
  \bibfield  {author} {\bibinfo {author} {\bibfnamefont {S.}~\bibnamefont
  {Pratt}}\ and\ \bibinfo {author} {\bibfnamefont {G.}~\bibnamefont
  {Torrieri}},\ }\bibfield  {title} {\bibinfo {title} {{Coupling Relativistic
  Viscous Hydrodynamics to Boltzmann Descriptions}},\ }\href
  {https://doi.org/10.1103/PhysRevC.82.044901} {\bibfield  {journal} {\bibinfo
  {journal} {Phys. Rev. C}\ }\textbf {\bibinfo {volume} {82}},\ \bibinfo
  {pages} {044901} (\bibinfo {year} {2010})},\ \Eprint
  {https://arxiv.org/abs/1003.0413} {arXiv:1003.0413 [nucl-th]} \BibitemShut
  {NoStop}%
\bibitem [{\citenamefont {Fotakis}\ \emph {et~al.}(2020)\citenamefont
  {Fotakis}, \citenamefont {Greif}, \citenamefont {Greiner}, \citenamefont
  {Denicol},\ and\ \citenamefont {Niemi}}]{Fotakis:2019nbq}%
  \BibitemOpen
  \bibfield  {author} {\bibinfo {author} {\bibfnamefont {J.~A.}\ \bibnamefont
  {Fotakis}}, \bibinfo {author} {\bibfnamefont {M.}~\bibnamefont {Greif}},
  \bibinfo {author} {\bibfnamefont {C.}~\bibnamefont {Greiner}}, \bibinfo
  {author} {\bibfnamefont {G.~S.}\ \bibnamefont {Denicol}},\ and\ \bibinfo
  {author} {\bibfnamefont {H.}~\bibnamefont {Niemi}},\ }\bibfield  {title}
  {\bibinfo {title} {{Diffusion processes involving multiple conserved charges:
  A study from kinetic theory and implications to the fluid-dynamical modeling
  of heavy ion collisions}},\ }\href
  {https://doi.org/10.1103/PhysRevD.101.076007} {\bibfield  {journal} {\bibinfo
   {journal} {Phys. Rev. D}\ }\textbf {\bibinfo {volume} {101}},\ \bibinfo
  {pages} {076007} (\bibinfo {year} {2020})},\ \Eprint
  {https://arxiv.org/abs/1912.09103} {arXiv:1912.09103 [hep-ph]} \BibitemShut
  {NoStop}%
\bibitem [{\citenamefont {Pratt}\ \emph {et~al.}(2011)\citenamefont {Pratt},
  \citenamefont {Schlichting},\ and\ \citenamefont {Gavin}}]{Pratt:2010zn}%
  \BibitemOpen
  \bibfield  {author} {\bibinfo {author} {\bibfnamefont {S.}~\bibnamefont
  {Pratt}}, \bibinfo {author} {\bibfnamefont {S.}~\bibnamefont {Schlichting}},\
  and\ \bibinfo {author} {\bibfnamefont {S.}~\bibnamefont {Gavin}},\ }\bibfield
   {title} {\bibinfo {title} {{Effects of Momentum Conservation and Flow on
  Angular Correlations at RHIC}},\ }\href
  {https://doi.org/10.1103/PhysRevC.84.024909} {\bibfield  {journal} {\bibinfo
  {journal} {Phys. Rev. C}\ }\textbf {\bibinfo {volume} {84}},\ \bibinfo
  {pages} {024909} (\bibinfo {year} {2011})},\ \Eprint
  {https://arxiv.org/abs/1011.6053} {arXiv:1011.6053 [nucl-th]} \BibitemShut
  {NoStop}%
\bibitem [{\citenamefont {Kuznietsov}\ \emph {et~al.}(2022)\citenamefont
  {Kuznietsov}, \citenamefont {Savchuk}, \citenamefont {Gorenstein},
  \citenamefont {Koch},\ and\ \citenamefont {Vovchenko}}]{Kuznietsov:2022pcn}%
  \BibitemOpen
  \bibfield  {author} {\bibinfo {author} {\bibfnamefont {V.~A.}\ \bibnamefont
  {Kuznietsov}}, \bibinfo {author} {\bibfnamefont {O.}~\bibnamefont {Savchuk}},
  \bibinfo {author} {\bibfnamefont {M.~I.}\ \bibnamefont {Gorenstein}},
  \bibinfo {author} {\bibfnamefont {V.}~\bibnamefont {Koch}},\ and\ \bibinfo
  {author} {\bibfnamefont {V.}~\bibnamefont {Vovchenko}},\ }\bibfield  {title}
  {\bibinfo {title} {{Critical point particle number fluctuations from
  molecular dynamics}},\ }\href {https://doi.org/10.1103/PhysRevC.105.044903}
  {\bibfield  {journal} {\bibinfo  {journal} {Phys. Rev. C}\ }\textbf {\bibinfo
  {volume} {105}},\ \bibinfo {pages} {044903} (\bibinfo {year} {2022})},\
  \Eprint {https://arxiv.org/abs/2201.08486} {arXiv:2201.08486 [hep-ph]}
  \BibitemShut {NoStop}%
\bibitem [{\citenamefont {Poberezhnyuk}\ \emph {et~al.}(2020)\citenamefont
  {Poberezhnyuk}, \citenamefont {Savchuk}, \citenamefont {Gorenstein},
  \citenamefont {Vovchenko}, \citenamefont {Taradiy}, \citenamefont {Begun},
  \citenamefont {Satarov}, \citenamefont {Steinheimer},\ and\ \citenamefont
  {Stoecker}}]{PhysRevC.102.024908}%
  \BibitemOpen
  \bibfield  {author} {\bibinfo {author} {\bibfnamefont {R.~V.}\ \bibnamefont
  {Poberezhnyuk}}, \bibinfo {author} {\bibfnamefont {O.}~\bibnamefont
  {Savchuk}}, \bibinfo {author} {\bibfnamefont {M.~I.}\ \bibnamefont
  {Gorenstein}}, \bibinfo {author} {\bibfnamefont {V.}~\bibnamefont
  {Vovchenko}}, \bibinfo {author} {\bibfnamefont {K.}~\bibnamefont {Taradiy}},
  \bibinfo {author} {\bibfnamefont {V.~V.}\ \bibnamefont {Begun}}, \bibinfo
  {author} {\bibfnamefont {L.}~\bibnamefont {Satarov}}, \bibinfo {author}
  {\bibfnamefont {J.}~\bibnamefont {Steinheimer}},\ and\ \bibinfo {author}
  {\bibfnamefont {H.}~\bibnamefont {Stoecker}},\ }\bibfield  {title} {\bibinfo
  {title} {Critical point fluctuations: Finite size and global charge
  conservation effects},\ }\href {https://doi.org/10.1103/PhysRevC.102.024908}
  {\bibfield  {journal} {\bibinfo  {journal} {Phys. Rev. C}\ }\textbf {\bibinfo
  {volume} {102}},\ \bibinfo {pages} {024908} (\bibinfo {year}
  {2020})}\BibitemShut {NoStop}%
\bibitem [{\citenamefont {Kuznietsov}\ \emph {et~al.}(2023)\citenamefont
  {Kuznietsov}, \citenamefont {Savchuk}, \citenamefont {Poberezhnyuk},
  \citenamefont {Vovchenko}, \citenamefont {Gorenstein},\ and\ \citenamefont
  {Stoecker}}]{Kuznietsov:2023iyu}%
  \BibitemOpen
  \bibfield  {author} {\bibinfo {author} {\bibfnamefont {V.~A.}\ \bibnamefont
  {Kuznietsov}}, \bibinfo {author} {\bibfnamefont {O.}~\bibnamefont {Savchuk}},
  \bibinfo {author} {\bibfnamefont {R.~V.}\ \bibnamefont {Poberezhnyuk}},
  \bibinfo {author} {\bibfnamefont {V.}~\bibnamefont {Vovchenko}}, \bibinfo
  {author} {\bibfnamefont {M.~I.}\ \bibnamefont {Gorenstein}},\ and\ \bibinfo
  {author} {\bibfnamefont {H.}~\bibnamefont {Stoecker}},\ }\bibfield  {title}
  {\bibinfo {title} {{Molecular dynamics analysis of particle number
  fluctuations in the mixed phase of a first-order phase transition}},\ }\href
  {https://doi.org/10.1103/PhysRevC.107.055206} {\bibfield  {journal} {\bibinfo
   {journal} {Phys. Rev. C}\ }\textbf {\bibinfo {volume} {107}},\ \bibinfo
  {pages} {055206} (\bibinfo {year} {2023})},\ \Eprint
  {https://arxiv.org/abs/2303.09193} {arXiv:2303.09193 [hep-ph]} \BibitemShut
  {NoStop}%
\bibitem [{\citenamefont {Sorensen}\ and\ \citenamefont
  {Koch}(2021)}]{Sorensen:2020ygf}%
  \BibitemOpen
  \bibfield  {author} {\bibinfo {author} {\bibfnamefont {A.}~\bibnamefont
  {Sorensen}}\ and\ \bibinfo {author} {\bibfnamefont {V.}~\bibnamefont
  {Koch}},\ }\bibfield  {title} {\bibinfo {title} {{Phase transitions and
  critical behavior in hadronic transport with a relativistic density
  functional equation of state}},\ }\href
  {https://doi.org/10.1103/PhysRevC.104.034904} {\bibfield  {journal} {\bibinfo
   {journal} {Phys. Rev. C}\ }\textbf {\bibinfo {volume} {104}},\ \bibinfo
  {pages} {034904} (\bibinfo {year} {2021})},\ \Eprint
  {https://arxiv.org/abs/2011.06635} {arXiv:2011.06635 [nucl-th]} \BibitemShut
  {NoStop}%
\bibitem [{\citenamefont {Pratt}(2017)}]{PhysRevC.96.044903}%
  \BibitemOpen
  \bibfield  {author} {\bibinfo {author} {\bibfnamefont {S.}~\bibnamefont
  {Pratt}},\ }\bibfield  {title} {\bibinfo {title} {Consistent implementation
  of non-zero-range terms into hydrodynamics},\ }\href
  {https://doi.org/10.1103/PhysRevC.96.044903} {\bibfield  {journal} {\bibinfo
  {journal} {Phys. Rev. C}\ }\textbf {\bibinfo {volume} {96}},\ \bibinfo
  {pages} {044903} (\bibinfo {year} {2017})}\BibitemShut {NoStop}%
\bibitem [{\citenamefont {Mukherjee}\ \emph {et~al.}(2015)\citenamefont
  {Mukherjee}, \citenamefont {Venugopalan},\ and\ \citenamefont
  {Yin}}]{Mukherjee_2015}%
  \BibitemOpen
  \bibfield  {author} {\bibinfo {author} {\bibfnamefont {S.}~\bibnamefont
  {Mukherjee}}, \bibinfo {author} {\bibfnamefont {R.}~\bibnamefont
  {Venugopalan}},\ and\ \bibinfo {author} {\bibfnamefont {Y.}~\bibnamefont
  {Yin}},\ }\bibfield  {title} {\bibinfo {title} {Real-time evolution of
  non-gaussian cumulants in the {QCD} critical regime},\ }\bibfield  {journal}
  {\bibinfo  {journal} {Physical Review C}\ }\textbf {\bibinfo {volume} {92}},\
  \href {https://doi.org/10.1103/physrevc.92.034912}
  {10.1103/physrevc.92.034912} (\bibinfo {year} {2015})\BibitemShut {NoStop}%
\bibitem [{\citenamefont {Bozek}\ and\ \citenamefont
  {Broniowski}(2013)}]{Bozek:2012is}%
  \BibitemOpen
  \bibfield  {author} {\bibinfo {author} {\bibfnamefont {P.}~\bibnamefont
  {Bozek}}\ and\ \bibinfo {author} {\bibfnamefont {W.}~\bibnamefont
  {Broniowski}},\ }\bibfield  {title} {\bibinfo {title} {{Charge balancing and
  the fall off of the ridge}},\ }\href
  {https://doi.org/10.1016/j.nuclphysa.2013.02.057} {\bibfield  {journal}
  {\bibinfo  {journal} {Nucl. Phys. A}\ }\textbf {\bibinfo {volume}
  {904-905}},\ \bibinfo {pages} {479c} (\bibinfo {year} {2013})},\ \Eprint
  {https://arxiv.org/abs/1210.4315} {arXiv:1210.4315 [nucl-th]} \BibitemShut
  {NoStop}%
\bibitem [{\citenamefont {Bozek}\ and\ \citenamefont
  {Broniowski}(2012)}]{Bozek:2012en}%
  \BibitemOpen
  \bibfield  {author} {\bibinfo {author} {\bibfnamefont {P.}~\bibnamefont
  {Bozek}}\ and\ \bibinfo {author} {\bibfnamefont {W.}~\bibnamefont
  {Broniowski}},\ }\bibfield  {title} {\bibinfo {title} {{Charge conservation
  and the shape of the ridge of two-particle correlations in relativistic
  heavy-ion collisions}},\ }\href
  {https://doi.org/10.1103/PhysRevLett.109.062301} {\bibfield  {journal}
  {\bibinfo  {journal} {Phys. Rev. Lett.}\ }\textbf {\bibinfo {volume} {109}},\
  \bibinfo {pages} {062301} (\bibinfo {year} {2012})},\ \Eprint
  {https://arxiv.org/abs/1204.3580} {arXiv:1204.3580 [nucl-th]} \BibitemShut
  {NoStop}%
\bibitem [{\citenamefont {Adam}\ \emph {et~al.}(2019)\citenamefont {Adam} \emph
  {et~al.}}]{PhysRevC.99.044918}%
  \BibitemOpen
  \bibfield  {author} {\bibinfo {author} {\bibfnamefont {J.}~\bibnamefont
  {Adam}} \emph {et~al.} (\bibinfo {collaboration} {STAR Collaboration}),\
  }\bibfield  {title} {\bibinfo {title} {Collision-energy dependence of
  ${p}_{t}$ correlations in au + au collisions at energies available at the bnl
  relativistic heavy ion collider},\ }\href
  {https://doi.org/10.1103/PhysRevC.99.044918} {\bibfield  {journal} {\bibinfo
  {journal} {Phys. Rev. C}\ }\textbf {\bibinfo {volume} {99}},\ \bibinfo
  {pages} {044918} (\bibinfo {year} {2019})}\BibitemShut {NoStop}%
\bibitem [{\citenamefont {Abelev}\ \emph {et~al.}(2009)\citenamefont {Abelev}
  \emph {et~al.}}]{PhysRevC.80.064912}%
  \BibitemOpen
  \bibfield  {author} {\bibinfo {author} {\bibfnamefont {B.~I.}\ \bibnamefont
  {Abelev}} \emph {et~al.} (\bibinfo {collaboration} {STAR}),\ }\bibfield
  {title} {\bibinfo {title} {{Long range rapidity correlations and jet
  production in high energy nuclear collisions}},\ }\href
  {https://doi.org/10.1103/PhysRevC.80.064912} {\bibfield  {journal} {\bibinfo
  {journal} {Phys. Rev. C}\ }\textbf {\bibinfo {volume} {80}},\ \bibinfo
  {pages} {064912} (\bibinfo {year} {2009})},\ \Eprint
  {https://arxiv.org/abs/0909.0191} {arXiv:0909.0191 [nucl-ex]} \BibitemShut
  {NoStop}%
\end{thebibliography}%
\end{document}